\documentclass[preprint,12pt]{elsarticle}

\usepackage{amssymb}
\usepackage{amsmath}
\usepackage[version=4]{mhchem}
\usepackage{hyperref}
\usepackage{float}
\usepackage{xcolor}
\usepackage{verbatim}

\journal{Computational Materials Science}

\begin{document}

\begin{frontmatter}

\title{Size focusing in core-shell precipitates: A phase-field study} %% Article title

\author{Soumya Mishra and T. A. Abinandanan} %% Author name

%% Author affiliation
\affiliation{organization={Department of Materials Engineering, Indian Institute of Science},   addressline={}, 
            city={Bengaluru},
            postcode={560012}, 
            state={Karnataka},
            country={India}} %Department and Organization

%% Abstract
\begin{abstract}

Due to their enhanced resistance to coarsening and/or creep, aluminium alloys with precipitates of two distinct phases in a core-shell morphology are of great contemporary interest. In this paper, we focus on the curious observation in two recent studies on Al-Sc-Li and Al-Yb-Li alloys that growth of the shell phase leads to a narrowing of the size distribution. We have studied this phenomenon, known as size focusing, using a theoretical framework (which extends Zener's theory of diffusional growth to a core-shell precipitate) and multi-precipitate simulations based on a phase field model. Our results yield key theoretical insights as well as conclusions with practical significance. (a) On the theoretical front, we show clearly that size focusing is a growth phenomenon: it ends when shell growth ends, and coarsening begins. (b) On the practical front, our results offer guidelines for designing alloys with narrower size distributions: size focusing is promoted in alloys with greater shell volume fractions and greater inter-precipitate spacing.

\end{abstract}

\begin{comment}
    
%%Graphical abstract
\begin{graphicalabstract}
\includegraphics{Figure_0.pdf}
\end{graphicalabstract}

%%Research highlights
\begin{highlights}
    \item Size focusing is a phenomenon in which size distribution becomes narrower, during the growth of a shell phase around pre-existing core precipitates. 
    \item Size focusing is essentially a growth phenomenon; it ends (and size \textit{defocusing} begins) when the growth of the shell phase ends and coarsening sets in.
    \item Size focusing is stronger in alloys with a higher shell volume fraction and lower core mobilities.
\end{highlights}

\end{comment}

%% Keywords

\end{frontmatter}

%% Add \usepackage{lineno} before \begin{document} and uncomment 
%% following line to enable line numbers
%% \linenumbers

%% main text

\section{Introduction}

Microstructures with ``composite'' precipitates, which are made up of a core of one precipitate phase enveloped in a shell of another phase,  were reported in an Al-Zr-Li alloy by Gayle and Vander Sande in 1984~\cite{gayle1984composite}. Since then, such core-shell microstructures have been the subject of numerous studies on Al-alloys \cite{gu1985influence, aydinol1994coarsening, karnesky2006effects, radmilovic2008monodisperse, vandalen2008creep, boothmorrison2011coarsening, vandalen2011effects, vandalen2011microstructural, monachon2011chemistry}; they continue to be of intense contemporary interest (see, for example, references \cite{deluca2018microstructure, dorin2022precipitation, schmid2022stabilization, ekaputra2022microstructure, leibner2022effect, leibner2022sc, liu2022correlation, wu2024er, mao2024strength, mondal2025understanding, jiang2025structurally, kong2025precipitation, li2025effects, baqeri2026nanoscale, yang2026microalloying, kumar2026high}) due to their role in enhanced resistance to coarsening \cite{boothmorrison2011coarsening, vandalen2011effects}, creep \cite{vandalen2011effects, vandalen2011microstructural}, and hydrogen-embrittlement \cite{jiang2025structurally}. 
 
In a landmark paper \cite{radmilovic2011highly}, Radmilovic et al. reported a curious finding: the narrowing of the precipitate size distribution in an Al-Li-Sc alloy. This phenomenon, known as size focusing, implies that the shell growth around smaller core precipitates is faster than that around larger core precipitates; as a result, smaller core precipitates are surrounded by thicker shells than larger core precipitates -- exactly as reported by Radmilovic et al. \cite{radmilovic2011highly}. A more dramatic version of this phenomenon (known as size focusing) was observed in a recent study by Chhotray and Gautam \cite{chhotray2026core} in an Al-Li-Yb alloy: the size distribution of core precipitates is bimodal, while that of core-shell (or composite) precipitates is unimodal. 

Size focusing is well known in the literature on wet-chemical synthesis for producing nearly monodisperse nanoparticles \cite{yin2005colloidal, qian2009size, jin2010size, peng1998kinetics}. Using a version of diffusion-controlled growth of particles from a liquid solution, Reiss developed a theory of size focusing \cite{reiss1951growth} that forms the basis for the use of repeated solute injection to produce nearly uniform-sized particles; see also References \cite{sugimoto1987preparation, clark2011focusing} for other theoretical approaches. 

However, precipitation in metallic alloys differs significantly from that in wet chemical synthesis. First, solute injection is clearly not possible in metallic alloys. Further, fast diffusion in the liquid phase (aided by continuous stirring in some cases) ensures a near-uniform solute concentration in the solution surrounding the particles; in contrast, the slow diffusion in metallic alloys occurring over longer distances leads to the solute concentration being almost always inhomogeneous within the matrix. Lastly, nanoparticles in liquid solutions have a near-zero volume fraction, whereas precipitates could have much higher volume fractions in metallic alloys .

In this paper, we present a theory of size focusing applicable to core-shell precipitates in solid alloys; our theory is based on extending Zener's model for diffusion-controlled growth of an isolated precipitate to the growth of an ensemble of precipitates. After validating this theory with model simulations with just two precipitates, we present our results from multi-precipitate simulations on the effect of shell volume fraction, core volume fraction, and core coarsening on size focusing.  

The paper is organised as follows: the phase-field model used for our simulations is described in Section~\ref{section_phase_field_model}. The theory of size focusing is presented in Section~\ref{section_theory_of_size_focusing}, followed by its validation using two-precipitate simulations. Results from extensive multi-precipitate simulations (on the influence of different factors on the extent and duration of size focusing) are presented in Section~\ref{section_multi_ppt_simulations}. The implications of these results are discussed in Section~\ref{section_discussion}, and key conclusions are summarised in Section~\ref{section_conclusions}.

\section{Phase Field Model}
\label{section_phase_field_model}

Our model is essentially a ternary version of the classical binary Cahn-Hilliard model \cite{cahn1958freeI, cahn1959freeII}. Since we draw extensively on the models used in studies of phase separation in ternary polymer blends \cite{huang1995phase} and ternary metallic alloys \cite{bhattacharyya2003study, makineni2017enhancing}, we only highlight the key steps. Further, the model presented here is in terms of dimensionless quantities; the composition variables are scaled in such a way that they are equal to the respective phase fractions. The non-dimensionalisation and scaling procedure are outlined in Section~J of the Supplementary Material.

Extending the classic study by Cahn and Hilliard \cite{cahn1958freeI, cahn1959freeII}, to a compositionally non-homogeneous ternary alloy, we write its total free energy as: 
\begin{equation}
	F = N_{v} \int_{V}\left[f_0\left(c_{A}, c_{B}, c_{c}\right) + \kappa_{A}\left(\nabla c_{A}\right)^2 + \kappa_{B}\left(\nabla c_{B}\right)^2 + \kappa_{C}\left(\nabla c_{C}\right)^2 \right] d V 
	\label{equation_1}
\end{equation}
where $N_{v}$ is the number of atoms per unit volume (assumed to be independent of composition), $f_0(c_{A}, c_{B}, c_{C})$ the (local) free energy per atom, and $\kappa_{i}$ is the gradient energy coefficient that penalises the composition gradient in species $i$. In this equation, $c_{i}$ is the mole fraction of species $i$, and $(c_{A}+c_{B}+c_{C})=1$ everywhere.

We use the following polynomial approximation to $f_0(c_{A}, c_{B}, c_{C})$ :
\begin{equation}
	f_0(c_{A}, c_{B}, c_{C}) =  \chi_{AB} c_{A}^2 c_{B}^2 + \chi_{BC} c_{B}^2 c_{C}^2 + \chi_{AC} c_{A}^2 c_{C}^2 + \chi_{ABC} c_{A}^2 c_{B}^2 c_{C}^2 \, ,
	\label{equation_2}
\end{equation}
where $\chi_{AB}$, $\chi_{BC}$, and $\chi_{AC}$ are the binary interaction energies between the respective pairs of species and $\chi_{ABC}$ is the ternary interaction energy between all three species. We emphasise that in Equation~\ref{equation_2} and all the following equations, $c_{A}$ has been substituted by $c_{A} = (1 - c_{B} - c_{C})$.

Defining the diffusion potentials $\mu_{B}^{\mathrm{eff}}$ and $\mu_{C}^{\mathrm{eff}}$ as the variational derivatives $\frac{\delta F}{\delta c_{B}}$ and $\frac{\delta F}{\delta c_{C}}$ respectively, we arrive at the following expressions:
\begin{align}
\mu_{B}^{\mathrm{eff}} = \frac{\delta F}{\delta c_{B}} & = \frac{\partial f_0}{\partial c_{B}}-2\left(\kappa_{B}+\kappa_{A}\right) \nabla^2 c_{B}-2 \kappa_{A} \nabla^2 c_{C} \nonumber \\
\mu_{C}^{\mathrm{eff}} = \frac{\delta F}{\delta c_{C}} & = \frac{\partial f_0}{\partial c_{C}}-2\left(\kappa_{A} + \kappa_{C}\right) \nabla^2 c_{C}-2 \kappa_{A} \nabla^2 c_{B} 
\label{equation_4}
\end{align}
where $\left(\partial f_0 / \partial c_{B}\right)$ and $\left(\partial f_0 / \partial c_{C}\right)$ are partial derivatives of $f_0$ in Equation~\ref{equation_2} after the substitution $c_{A}=(1 - c_{B} - c_{C}$).

Finally, the time-evolution of the composition fields $c_{B}$ and $c_{C}$ is governed by (see Section~H in Supplementary Material for details) : 
\begin{align*}
	\frac{\partial c_{B}}{\partial t} & = M_{B B}\left[\nabla^2\left(\frac{\partial f_0}{\partial c_{B}}\right)-2\kappa_{AB} \nabla^{4} c_{B}-2 \kappa_{A} \nabla^{4} c_{C}\right] \\
	& + M_{B C}\left[\nabla^2\left(\frac{\partial f_0}{\partial c_{C}}\right)-2\kappa_{AC} \nabla^{4} c_{C}-2 \kappa_{A} \nabla^{4} c_{B}\right] 
\end{align*}
\begin{align}
	\frac{\partial c_{C}}{\partial t} & = M_{C C}\left[\nabla^2\left(\frac{\partial f_0}{\partial c_{C}}\right)-2\kappa_{AC} \nabla^{4} c_{C}-2 \kappa_{A} \nabla^{4} c_{B}\right] \nonumber \\
	& + M_{B C}\left[\nabla^2\left(\frac{\partial f_0}{\partial c_{B}}\right)-2\kappa_{AB} \nabla^{4} c_{B}-2 \kappa_{A} \nabla^{4} c_{C}\right] 
	\label{equation_5}
\end{align}
where $\kappa_{AB} = \kappa_{A} + \kappa_{B}$ and $\kappa_{AC} = \kappa_{A} + \kappa_{C}$.

A numerical solution to these equations is used for studying microstructural evolution; the semi-implicit method used in our study is the same as that in Reference~\cite{bhattacharyya2003study}  (which, in turn, extends the method of Chen and Shen \cite{chen1998applications} for binary systems). We have used the CUDA C programming language to implement this method with GPU parallelisation. The numerical implementation is described in detail in Section~I in the Supplementary Material.

In phase field simulations, the microstructure at a given time $t$ is in the form of the compositions $(c_B(\textbf{r}), c_C(\textbf{r}))$ at position $\textbf{r}$. At each grid point $\textbf{r}$, we select the species with the highest concentration among the three, and assign that point to its corresponding phase (and colour using the palette in Figure~G.1 in the Supplementary Material). For quantitative analysis of these microstructures, we use the Hoshen-Kopelman algorithm \cite{hoshen1976percolation} for each precipitate phase; from the output of this step, we compute for each precipitate its area $A_i = N\Delta x \Delta y$ and its equivalent radius $r_i = A_i/ \pi$.

Table~\ref{table_size_focusing_simulation_parameters} lists the model parameters chosen to obtain interfacial energies which obey the wetting condition: $\sigma_{\alpha\beta} > (\sigma_{\beta\gamma} + \sigma_{\alpha\gamma})$. [We note here the role of $\chi_{ABC}$ in Equation~\ref{equation_1}: our choice of $\chi_{ABC}$ ensure a low level of segregation ($c_{B} \leq 0.04$) of species $B$ (say) at the interface between the $A$-rich $\alpha$  and $C$-rich $\gamma$ phases.]

\begin{table}[H]
    \centering
    \begin{tabular}{|c|c|c|c|c|c|c|}
        \hline
        $(\chi_{AB}$, $\kappa_{AB}$)
        & ($\chi_{BC}$, $\kappa_{BC}$)
        & ($\chi_{AC}$, $\kappa_{AC}$)
        & $\chi_{ABC}$ 
        & $\sigma_{\alpha\beta}$
        & $\sigma_{\beta\gamma}$
        & $\sigma_{\alpha\gamma}$\\ 
        \hline        
        (3, 4.88) & (1, 2.45) & (1, 2.45) & 50 & 1.065 & 0.5 & 0.5 \\
        \hline
    \end{tabular} 
    \caption{Model parameters and interfacial energies}
    \label{table_size_focusing_simulation_parameters}
\end{table}

\section{Theory of size focusing}
\label{section_theory_of_size_focusing}

We consider a ternary alloy in which precipitation of a $B$-rich $\beta$ core phase (in a previous heat treatment) had proceeded well into the coarsening regime with little or no supersaturation of $B$ left in the matrix $\alpha$ phase; thus, at the beginning of the second heat treatment (for shell formation), the matrix has a near uniform composition of the second solute species $C$ (see the composition profile in Figure~A.1 (a) in the Supplementary Material). During the early stages of this second heat treatment, the $C$-rich shell phase $\gamma$ spontaneously forms and grows over the $\beta$ (core) precipitates [see Figure~A.1 (b) for the composition profile after the formation of a thin shell layer]; we assume, of course, that the relevant interfacial energies obey the wetting condition ($\sigma_{\alpha\beta} > \sigma_{\alpha\gamma} + \sigma_{\beta\gamma}$). 

Since shell growth around each precipitate $i$ proceeds in a matrix with constant supersaturation $c_{C}^{0}$ of species $C$, Zener's theory of precipitate growth is applicable. A key result of this theory is that the square of the radius  $r_i$ of each precipitate increases linearly with time $t$: 
\begin{equation}
r_i^{2} \thinspace = r_{i, 0}^2 + K \thinspace t \thinspace ,
\label{equation_r_vs_t_Zener_paper}
\end{equation}
where $r_{i,0}$ is the radius of the core precipitate $i$ $t = 0$ of the second heat treatment, $K$ is the growth coefficient which increases linearly with solute diffusivity $D$ (or, effective $D$ in alloys in which the diffusivity not constant \cite{mukherjee2009phase, mukherjee2010precipitate}), and monotonically with supersaturation $c_{C}^{0}$. [We note here that Zener's theory assumes that the role of interface curvature during growth in two or three dimensions (2D or 3D) is negligible.] 

The above relation implies the following expression for the growth rate $dr/dt$: 
\begin{equation}
\dfrac{dr_i}{dt} = \dfrac{K}{2 \thinspace r_i} \thinspace .
\label{equation_dr_dt_vs_r_Zener}
\end{equation}
This inverse dependence of ($dr_{i}/dt$) on precipitate radius ($r_{i}$) implies that the radius of a smaller precipitate (say) grows faster than that of a larger one; this forms the physical basis for size focusing. In order to track the rate of narrowing of the size distribution, we use the following definition of $\sigma^{s}$, the scaled standard deviation for an ensemble of $n$ precipitates: 
\begin{equation}
  \sigma^s = \frac{\sigma}{\overline{r}} \thickspace ,
  \label{equation_scaled_standard_deviation}
\end{equation}
where $\overline{r}$ and $\sigma$ are, respectively, the average and standard deviation for the ensemble: $\overline{r} = (1/n) \sum_{i=1}^ {n} r_{i} $ and $\sigma^{2} = (1/n) \sum_{i=1}^{n} (r_i - \overline{r})^2 $. Size focusing may now be defined as a process during which $\sigma^{s}$ decreases with time (or $(d \sigma^{s} / dt )$ is negative). 

Substituting for $\sigma$ and $\overline{r}$ in Equation~\ref{equation_scaled_standard_deviation} and differentiating the result with respect to time, we get the following (equivalent) expressions for  $\left[\, d \sigma^{s} / dt \,\right] $ (see Section~B in the Supplementary Material for details): 
\begin{subequations} % For naming 6a, 6b etc
\begin{align}
 \dfrac{d\sigma^s}{dt} \thinspace & = \thinspace \dfrac{-K}{2 \thinspace \sigma^s \thinspace \overline{r}^3 \thinspace H} \left[\overline{r^2} - H \thinspace \overline{r}\right] \thinspace 
\label{equation_dsigma_s_dt_multi_ppt_a} \\ 
 & = \thinspace \dfrac{-K}{2 \thinspace \sigma^s \thinspace \overline{r}} \left[\dfrac{1 \thinspace + \thinspace {\sigma^s}^2}{H} - \dfrac{1}{\overline{r}}\right] \thinspace , 
\label{equation_dsigma_s_dt_multi_ppt_b}
\end{align}
\end{subequations}
where $H = n/ (\sum_{i=1}^ {n} \left[ \, 1 / r_{i} \, \right]$) is the harmonic mean of the ensemble and $\overline{r^{2}} = (1/n) \sum_{i=1}^ {n} r_{i}^{2}$. 
Now, since $\overline{r} > H$ and $\overline{r^2} > \overline{r}^2$ for any ensemble of $n$ precipitates with $r_i>0$, we have $\overline{r} \thinspace \overline{r^2} > H \thinspace \overline{r}^2$ and therefore, $\overline{r^2} > H \thinspace \overline{r}$. The term within the square brackets, as well as the other quantities on the right-hand side of Equation~\ref{equation_dsigma_s_dt_multi_ppt_a}, are positive; therefore, $(d\sigma^{s} / dt)$ is negative, which implies size focusing or narrowing of size distribution.  

For a system with only two precipitates of radii $r_{l}$ and $r_{s}$ (with $r_{l} > r_{s}$), we have $\overline{r} = (r_l + r_s) / 2$, $\sigma = (r_l - r_s) / 2$, and $\sigma^s = (r_l - r_s) / (r_l + r_s)$, and the time evolution of $\sigma^s$ may be simplified to yield the following equivalent expressions: 
\begin{equation}
 \dfrac{d\sigma^s}{dt} \thinspace = \thinspace \dfrac{- \thinspace K \thinspace \sigma^s}{r_l \thinspace r_s} \thinspace = \thinspace \dfrac{-\thinspace K \thinspace \sigma^s}{\overline{r}^2 \thinspace (1 - {\sigma^s}^2)} \thinspace.
\label{equation_dsigma_s_dt_two_ppt}
\end{equation} 

If the precipitates are at infinite separation, size focusing would proceed until all the size differences vanish and $\sigma^{s}$ reaches zero. In alloys with a finite volume fraction (and hence finite inter-precipitate separation), this equation is valid only during the `classical' growth phase, during which diffusion fields around precipitates do not overlap. We may note that even after the overlap of diffusion fields, size focusing could still proceed, albeit at a lower rate, until the supersaturation is exhausted. [An estimate for the duration ($\tau$) of the classical growth regime is given in Section~F of the Supplementary Material.]

Assuming negligible diffusion of the core-forming species (i.e., the radius of each core precipitate does not change with time), it can be shown that shell width ($w_{\gamma}$) and core radius ($r_{\beta}$) are anti-correlated. Defining $r_{\beta}$ and $r_{\beta\gamma}$ respectively as the radius of the core and composite precipitate, the shell width $w_{\gamma} = r_{\beta\gamma} - r_{\beta}$. Since $r_{\beta\gamma} = \sqrt{r_{\beta}^2 + K \thinspace t}$ from Equation~\ref{equation_r_vs_t_Zener_paper}, we have:
\begin{equation}
  w_{\gamma} = \sqrt{r_{\beta}^2 + K \thinspace t} - r_{\beta} \thinspace. 
\label{equation_r_beta_w_gamma} 
\end{equation}
Differentiation with respect to $r_{\beta}$ yields:
\begin{align}
    \dfrac{dw_{\gamma}}{dr_{\beta}} = \dfrac{r_{\beta}}{\sqrt{r_{\beta}^2 + K \thinspace t}} - 1 = \dfrac{r_{\beta} - r_{\beta\gamma}}{r_{\beta\gamma}} < 0 \thinspace, 
\label{equation_derivative_r_beta_w_gamma} 
\end{align}
since $r_{\beta} < r_{\beta\gamma}$.

\subsection{Two precipitate simulation: comparison with theory}
\label{section_validation_with_theory_two_ppt}

We have simulated shell growth around two precipitates to validate the result in Equation~\ref{equation_dsigma_s_dt_two_ppt} and to elucidate the key features of size focusing. The 2D simulation uses a box of size 512 $\times$ 512 with two core precipitates with $c_{B} = 1$: the smaller one with radius $r_s \approx 24.5$ at the centre and the larger one with $r_l \approx 49$ at the corner so that $({r_l}/{r_s}) \approx 2$. With periodic boundary conditions, this yields a core volume fraction ($v_{\beta}$) of $\approx 0.04$. The matrix composition ($(c_{B}, c_{C})$) is chosen such that the overall alloy composition is set at $(c_{B}^{0}, c_{C}^{0}) = (0.04, 0.12)$ (the same as that used in the multi-particle simulations in Section~\ref{section_multi_ppt_simulations}). The mobility of the core-forming species is set at a far lower value ($M_{BB}=0.01)$ than that of the shell-forming species ($M_{CC}=1.0$).   

\begin{figure}[H]
\centering
\includegraphics{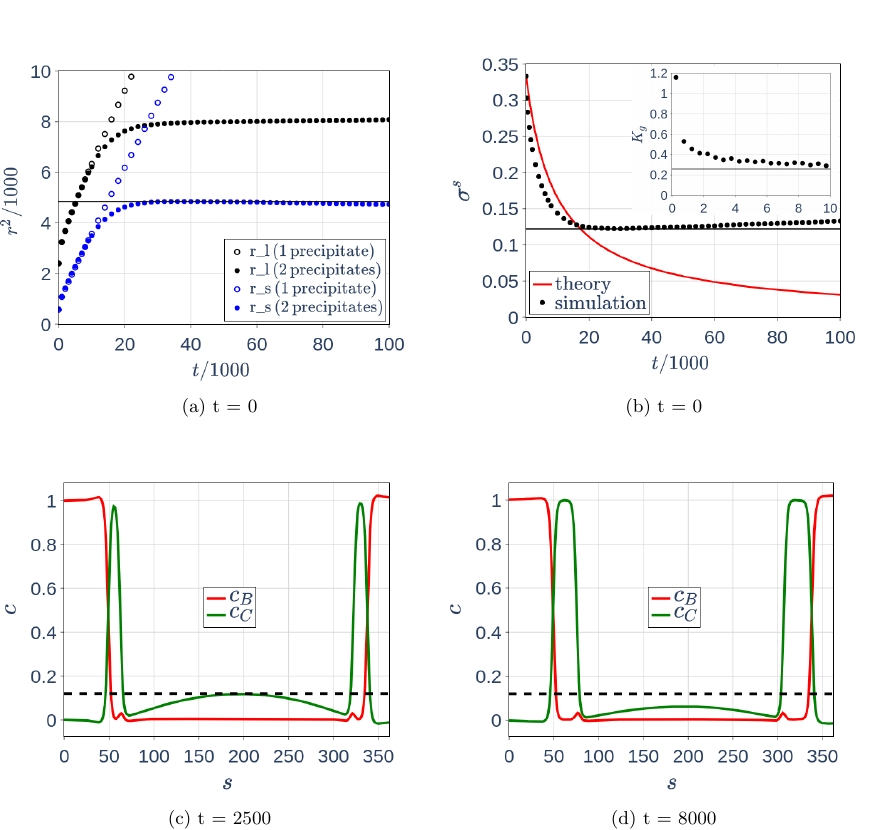}
    \caption{(a) Time dependence of precipitate radii for the alloy with $(c_{B}^{0}, c_{C}^{0}) = (0.04, 0.12)$ from two-precipitate simulations (filled dots in blue and black) and from two (separate) simulations with a single large (open black dots) and single small (open blue dots) precipitates); the resulting time evolution of scaled standard deviation ($\sigma^s$) is shown in (b). (c) and (d) Composition profiles from the centre of the large precipitate ($s = 0$) to that of the small precipitate ($s = \dfrac{512\sqrt{2}}{2}\approx 362$) at $t = 2500$ and $t = 8000$, indicating classical and non-classical growth, respectively. For clarity, we plot only $c_{B}$ and $c_{C}$ \textit{vs.} distance ($s$).}
    \label{figure_size_focusing_two_ppt_classical_non_classical_growth}
\end{figure}

The shell phase $\gamma$ wets the interface around the core $\beta$ precipitates and grows around them.  In Figure~\ref{figure_size_focusing_two_ppt_classical_non_classical_growth}(a), we show the evolution of $r^2$ as a function of time $t$ for the large and small precipitates from the two-precipitate simulations (filled dots in black and blue, respectively); also shown in the same figure are data from two separate simulations, one with a single large precipitate (open black dots) and the other with a single small one (open blue dots). The data for the growth of (isolated) single precipitates fall on parallel lines after an initial transient; in other words, they follow (to a good approximation) Equation~\ref{equation_dr_dt_vs_r_Zener} for Zener dynamics. The faster growth during the transient is clearly due to the $\gamma$ phase wetting the interface around the $\beta$ precipitate. 

More importantly, the data from single- and two-precipitate simulations show three clear stages: (a) a strong overlap during the initial stage, indicating that growth of both large and small precipitates (in the two-precipitate simulations) is the same as that of their isolated counterparts; (b) beyond $t \sim 2500$ both precipitates grow, but at ever slower rates than their isolated counterparts; and beyond $t \sim 25,000$, the small precipitate shrinks, while the large precipitate continues to grow (albeit much more slowly than earlier). These three stages represent, respectively, (a) classical growth (wherein the diffusion fields around precipitates have little or now overlap, see Figure~\ref{figure_size_focusing_two_ppt_classical_non_classical_growth}(c)), (b) non-classical growth after the diffusional overlap ( see Figure~\ref{figure_size_focusing_two_ppt_classical_non_classical_growth}(d)), and finally, (c) coarsening wherein the larger precipitate grows at the expense of the smaller one since the precipitates compete for solute after their growth has consumed almost all the supersaturation. 

The growth coefficient $K$ may be computed at different times using the $r^2$ \textit{vs.} $t$ data in  Figure~\ref{figure_size_focusing_two_ppt_classical_non_classical_growth}(a) from single precipitate simulations; the result is shown in the inset in  Figure~\ref{figure_size_focusing_two_ppt_classical_non_classical_growth}(b). Since shell formation is driven by spontaneous wetting, the (instantaneous) growth is much faster during the early stages; after the shell is fully formed, the growth rate keeps dropping to eventually reach a size- and time-independent value of $K = 0.26$ beyond $t \sim 80,000$. This value may now be used for computing $\sigma^{s}$ as a function of time for two precipitates used in simulation by numerically integrating Equation~\ref{equation_dsigma_s_dt_two_ppt}. The red curve in Figure~\ref{figure_size_focusing_two_ppt_classical_non_classical_growth}(b) shows the time evolution of $\sigma^{s}$ from the theory and simulation. 

First, $\sigma^{s}$ shows a steep drop right from the beginning (which is clearly due to the size-focusing effect), reaches a minimum at $t \sim 25000$ and shows a steady (but much slower) increase beyond this time. Thus, size focusing (the drop in $\sigma^{s}$) spans the period that includes both classical growth until $t \sim 2500$ (see Figure~\ref{figure_size_focusing_two_ppt_classical_non_classical_growth}(c)) and non-classical growth with ever-decreasing supersaturation during $2500 \leq t \leq 25000$ (see Figure~\ref{figure_size_focusing_two_ppt_classical_non_classical_growth}(d)); in particular, it ends when growth ends, after which the opposite effect of size defocusing sets in due to coarsening, a competitive process wherein the growth of the larger precipitate is accompanied by a shrinking of the smaller precipitate.   

Since the theory is based only on classical growth, it predicts a steady (but ever-decreasing) drop in $\sigma^{s}$ at all times, depicted using the red curve in Figure~\ref{figure_size_focusing_two_ppt_classical_non_classical_growth}(b). In this figure, the decrease in $\sigma^{s}$ predicted by the theory is slower than that from the simulation. This is primarily because the precipitates in the simulation show a much faster growth during the early stages (as discussed above and shown in the inset in Figure~\ref{figure_size_focusing_two_ppt_classical_non_classical_growth}(b)). 

The factors (such as shell volume fraction, core volume fraction and core mobility) that affect size focusing in two-precipitate simulations are presented in Section~C in the Supplementary Material (see also \cite{mishra2025coreshell}). 

We further show in Section~C.5 (in Supplementary Material) that, as long as the wetting condition is satisfied: $S_{\gamma} = \sigma_{\alpha\beta} - (\sigma_{\beta\gamma} + \sigma_{\alpha\gamma}) > 0$, the relative values of the three interfacial energies do not have a significant effect in determining the strength or duration of size focusing. This can indeed be seen in Figure~C.5 in Supplementary Material, where $\sigma^s$ \textit{vs} $t$ behaviour is near-identical for four systems with different combinations of interfacial energies.

\section{Results}
\label{section_multi_ppt_simulations}

Our multi-precipitate simulations mimic the two-heat treatments used by Radmilovic et al. \cite{radmilovic2011highly, radmilovic2008monodisperse}: the precipitation of the core ($\beta$) phase is simulated first, followed by a second simulation for the formation of the shell ($\gamma$) phase. The first simulation begins with randomly placed core ($\beta$) nuclei (with size above critical radius) in a binary (A-B) alloy with composition $(c_{B}^{0} = 0.04)$, and allow them to grow and coarsen till only 50 \% of them remain; we use a binary system for this step to avoid spontaneous shell formation that occurs in a ternary alloy even during the growth of the core precipitates. This microstructure is shown in Figure~\ref{figure_size_focusing_multi_ppt_microstructures}(a).

In the second simulation, we introduce the third species (C) in the matrix such that the overall alloy composition becomes equal to $c_{C}^{0}$. Unlike the two-precipitate simulations, we use equal mobilities for $B$ and $C$ ($M_{BB}$ and $M_{CC}$ = 1); this implies that coarsening of $\beta$ precipitates would continue even during the formation and growth of the shell phase (a system with lower $M_{BB}$ is considered in Section~\ref{section_multi_ppt_core_coarsening}). The simulation parameters are presented in Table~\ref{table_size_focusing_multi_ppt_simulations}; the values of grid-spacing ($\Delta x$ and $\Delta y$) and time-step ($\Delta t$) have been chosen in such a way that (a) time-integration is numerically stable and (b) all the interfaces are well-resolved (atleast six grid-points in the interface region).

All the microstructures that are relevant to the results in this section can be found in Section~D in the Supplementary Material. The statistical analysis presented in the following section is computed using the data from four independent simulations. 

\begin{table}[H]
    \centering
    \begin{tabular}{|c|c|}
        \hline
        ($\Delta x$, $\Delta y$) & (1.0, 1.0) \\
        \hline 
        $\Delta t$ & 0.25 \\
        \hline 
        ($L_x$, $L_y$) & (4096, 4096) \\
        \hline 
        ($c_{B}^{0}$, $c_{C}^{0}$) & (0.04, 0.12) \\
        \hline 
        ($M_{BB}$, $M_{CC}$, $M_{BC}$) & (1.0, 1.0, 0.0) \\
        \hline  
    \end{tabular}
    \caption{Simulation parameters}
    \label{table_size_focusing_multi_ppt_simulations}
\end{table}

In the following subsections, where we compare size focusing in different alloys or systems, we analyse our results in terms of strength of size focusing $s_{\text{sf}}$ defined as the difference between $\sigma^s$ at the beginning of the simulation and at the minimum ($s_{\text{sf}} = \sigma_{0}^{s} - \sigma_{\text{min}}^{s}$) and (b) duration of size focusing, $t_{\text{sf}}$, defined as the time at which $\sigma^{s}=\sigma_{\text{min}}^{s}$.

\subsection{Size focusing in the alloy with \texorpdfstring{$(c_{B}^{0}, c_{C}^{0}) = (0.04, 0.12)$}{(cB0, cC0) = (0.04, 0.12)}}
\label{section_multi_ppt_alloy_cb0_4_cc0_12}

\begin{figure}[H]
\centering
\includegraphics{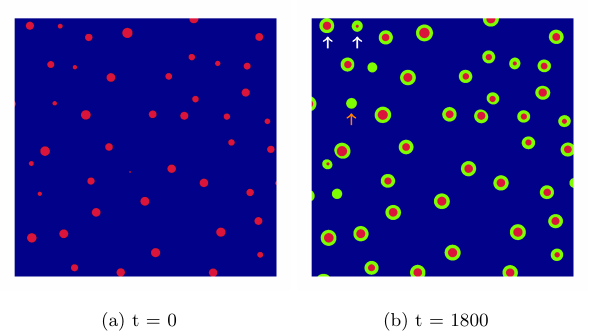}
\caption{Microstructural evolution during the second treatment in the alloy with $(c_{B}^{0}, c_{C}^{0}) = (0.04, 0.12)$. For clarity, only a 1024 x 1024 section (from a simulation box with 4096 x 4096 grid points) is shown. In (b), the precipitates marked using white arrows indicate the anti-correlation between core radius and shell width, and the precipitate marked using the orange arrow is an isolated shell ($\gamma$) left behind by dissolving core precipitates.}
\label{figure_size_focusing_multi_ppt_microstructures}
\end{figure}

The microstructure at $t = 1800$ (during the second treatment) is shown in Figure~\ref{figure_size_focusing_multi_ppt_microstructures}(b); the white arrows indicate two precipitates showing clearly that the smaller of the two is surrounded by a thicker shell. The precipitate marked with the orange arrow is an example of an isolated shell ($\gamma$) left behind by core precipitates that dissolved due to coarsening. In the rest of the paper, the results are presented in the form of statistical descriptors of the microstructure.

\begin{figure}[H]
\centering
\includegraphics{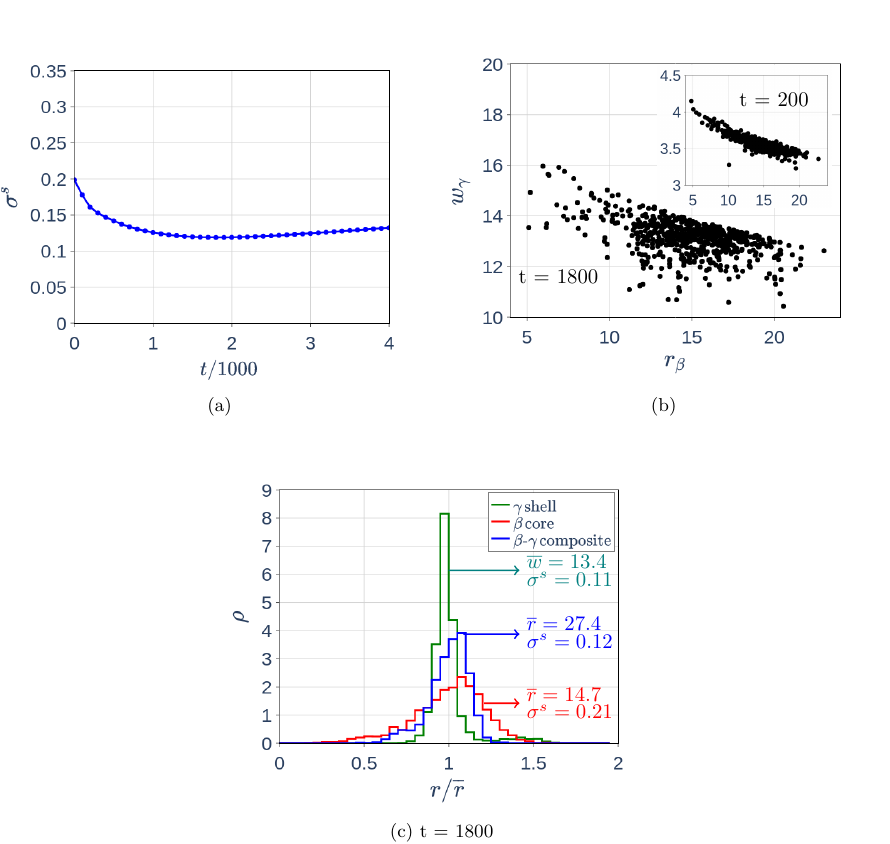}
    \caption{(a) Time evolution of the scaled standard deviation ($\sigma^s$) for all precipitates for the alloy $(c_{B}^{0}, c_{C}^{0}) = (0.04, 0.12)$; (b) inverse relationship between the shell width ($t_{\gamma}$) and core radius ($r_{\beta}$) at $t = 1800$; the inset shows a lower scatter at an earlier time $t = 200$ (c) scaled size distributions of core, shell, and composite precipitates at $t = 1800$, showing average radii (widths) and scaled standard deviations of each phase ($n = 2636$ precipitates).}
    \label{figure_size_focusing_multi_ppt_std_dev_core_shell_rad_size_dist}
\end{figure}

Figure~\ref{figure_size_focusing_multi_ppt_std_dev_core_shell_rad_size_dist}(a) presents the $\sigma^s$ \textit{vs.} t curve from multi-precipitate simulation. Just as in the two-precipitate simulations (Figure~\ref{figure_size_focusing_two_ppt_classical_non_classical_growth}(b)), (a) $\sigma^s$ \textit{vs.} t curve shows a sharp initial drop and reaches its minimum (by $t \sim 1800$), beyond which, (b) it shows an increase signifying size defocusing due to coarsening. The inverse correlation between shell width and core radius is demonstrated in Figure~\ref{figure_size_focusing_multi_ppt_std_dev_core_shell_rad_size_dist}(b), which is a scatter plot of shell width (on the y-axis) and core radius (on the x-axis) of precipitates from one simulation at $t = 1800$, where $\sigma^s$ is at its minimum. The scatter in the data reflects the role of the local environment (i.e.,  overlap of diffusion fields from neighbouring precipitates); this scatter is far lower at an earlier time $t = 200$ (see the inset in Figure~\ref{figure_size_focusing_multi_ppt_std_dev_core_shell_rad_size_dist}(b)), where the overlap of diffusion fields is less significant. 

Figure~\ref{figure_size_focusing_multi_ppt_std_dev_core_shell_rad_size_dist}(c) shows the scaled distributions for core radius, shell width and composite ($\beta\text{-}\gamma$) radius at $t = 1800$; the $\beta\text{-}\gamma$ precipitate size distribution is narrower ($\sigma^s = 0.12$) than that of core precipitates ($\sigma^s = 0.21$). Interestingly, the size distribution for shell width ($\sigma^s = 0.11$) shows a small secondary peak at higher values; a detailed analysis reveals that the dominant first peak comprises of shells of composite $\beta\text{-}\gamma$ precipitate, and the small secondary peak at higher $w$ values corresponds to isolated $\gamma$ precipitates left behind by the core precipitates that have dissolved away due to coarsening. [For such precipitates, the radius is taken as the shell width.]
 
In the experimental findings by Radmilovic et al. \cite{radmilovic2011highly}, the Full Width at Half Maximum (FWHM) values are similar for the size distributions for core radius and shell width  (0.24). In contrast, our results show that size focusing can be achieved even when $\sigma^s$ for the shell width is much lower than that of the core radius.

We note that the duration of size focusing ($t_\mathrm{sf}$) in multi-precipitate simulations (1800) is lower than that in the two-precipitate simulations (25,000). This difference is due to a combination of three factors: (a) inter-precipitate distances are significantly different ($d \sim 160$ in the former and $d \sim 362$ in the latter); (b) core coarsening is quite significant in the former (for example, by $t = 1800$, 145 ($\approx$ 5 \%) core precipitates had dissolved) and (c) due to differences in local environments, some precipitates may enter coarsening sooner due to their smaller-than-average separation (see Figure~\ref{figure_size_focusing_multi_ppt_std_dev_core_shell_rad_size_dist}(b)).

\subsection{Role of shell volume fraction}
\label{section_multi_ppt_shell_vol_fr}

\begin{figure}[H]
\centering
\includegraphics[scale = 0.5]{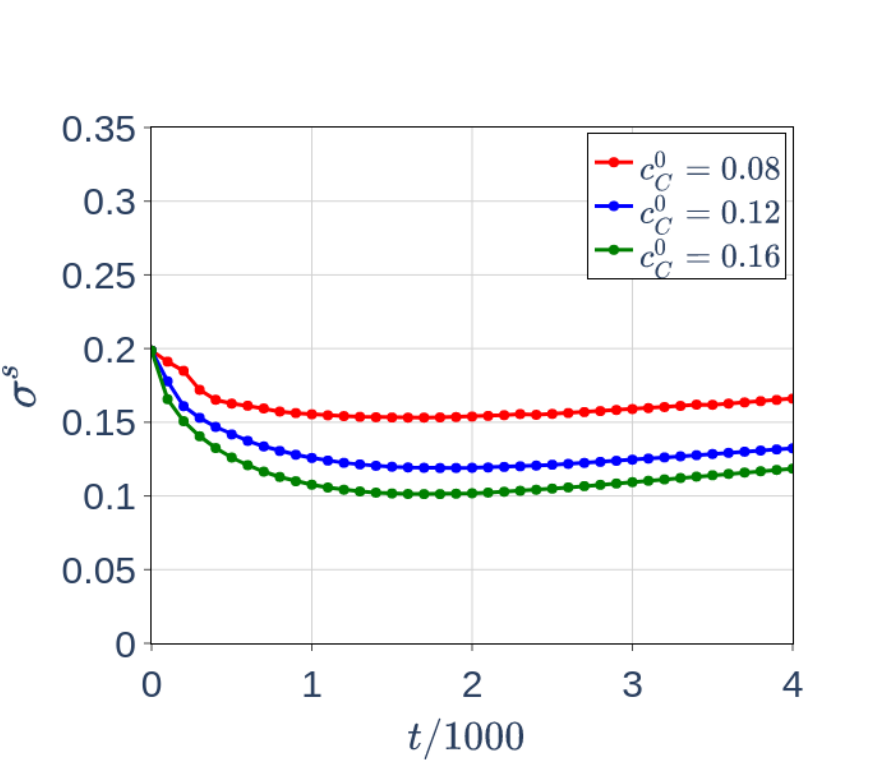}
\caption{Time evolution of scaled standard deviation ($\sigma^s$) for three alloys with different shell volume fraction $c_{C}^{0}$ = 0.08, 0.12, 0.16, but with constant core volume fraction $c_{B}^{0}$ = 0.04.}
    \label{figure_size_focusing_multi_ppt_shell_vol_fr_scaled_std_dev}
\end{figure}

Figure~\ref{figure_size_focusing_multi_ppt_shell_vol_fr_scaled_std_dev}(a) shows the time evolution of $\sigma^s$ in three alloys with different shell volume fractions ($c_{C}^{0} = 0.08, 0.12, 0.16$) but with the same core volume fraction ($c_{B}^{0} = 0.04$). While the initial microstructures are identical (with $\sigma^s =0.199)$) for all the three alloys, the curve for $c_{C}^{0} = 0.16$ (in blue) is steeper and $\sigma^s_{\text{min}}$ is lower than those for the other two alloys, indicating that size focusing is stronger in alloys with higher supersaturation in the shell-forming species ($c_{C}^{0}$); this is due to faster shell growth (or higher $K$ in Equation~\ref{equation_r_vs_t_Zener_paper}) in alloys with higher supersaturation. As we showed in Section~\ref{section_theory_of_size_focusing}, higher $K$ leads to a sharper drop in $\sigma^s$ (see Equation~\ref{equation_dsigma_s_dt_multi_ppt_a}).

On the other hand, the duration of size focusing in ~\ref{figure_size_focusing_multi_ppt_shell_vol_fr_scaled_std_dev} and Table~\ref{table_size_focusing_multi_ppt_shell_vol_fr} is the same for all three alloys. This is not surprising because it depends only on inter-precipitate spacing, which is identical in all three alloys. [These results are broadly in line with two-precipitate simulations presented in Section~C.2 in the Supplementary Material.]

\begin{table}[H]
\centering
    \begin{tabular}{|c|ccc|}
    \hline
    $c_{C}^{0}$
        & $t_\mathrm{sf}$ 
        & $\sigma^s_\mathrm{min}$
        & $s_\mathrm{sf}$ \\
    \hline
        0.08 & 1600 & 0.153 & 0.045 \\
        0.12 & 1800 & 0.119 & 0.08 \\
        0.16 & 1800 & 0.101 & 0.097 \\
    \hline    
    \end{tabular}
    \caption{Metrics of size focusing estimates ($\sigma^s_\mathrm{min}$ and $s_\mathrm{sf}$) for the three alloys shown in Figure~\ref{figure_size_focusing_multi_ppt_shell_vol_fr_scaled_std_dev}; the differences in $t_\mathrm{sf}$ are deemed to be insignificant due to the shallowness of the minima in the $\sigma^s$ \textit{vs.} $t$ curves in Figure~\ref{figure_size_focusing_multi_ppt_shell_vol_fr_scaled_std_dev}.}
    \label{table_size_focusing_multi_ppt_shell_vol_fr}
\end{table}

\subsection{Role of core volume fraction}
\label{section_size_focusing_multi_ppt_core_vol_fr}

\begin{figure}[H]
\centering
\includegraphics{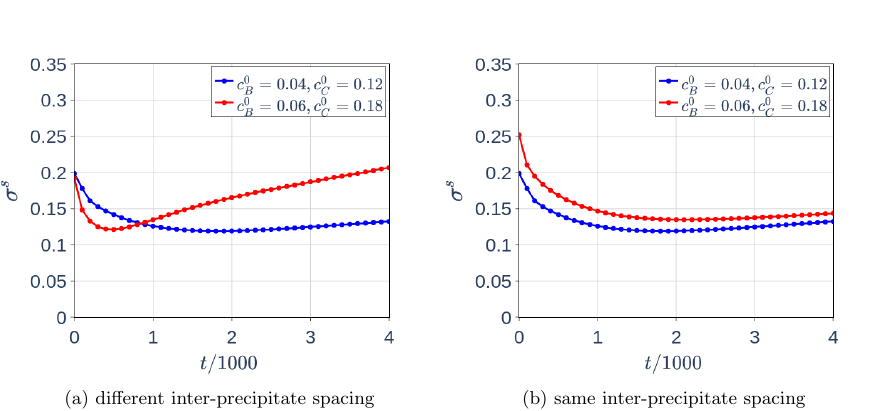}
    \caption{Time evolution of scaled standard deviation ($\sigma^s$)
 of all precipitates for two alloys with identical core-to-shell volume fraction ratio (1:3) but with different core volume fractions ($c_{B}^{0} = 0.04, 0.06$)}
    \label{figure_size_focusing_core_vol_fr}
\end{figure}

We now consider two alloys $(c_{B}^{0}, c_{C}^{0}) = (0.04, 0.12)$ and $(c_{B}^{0}, c_{C}^{0}) = (0.06, 0.18)$ with different core volume fractions but identical core-to-shell volume ratios (1:3). We present the results in two different ways: in the first, the first simulation for both the alloys is halted after half of the core precipitates have dissolved due to coarsening. Figure~\ref{figure_size_focusing_core_vol_fr}~(a) shows the time evolution of $\sigma^s$ in these two alloys. Starting from nearly the same $\sigma^s$ values, the red curve, for the more concentrated alloy with $(c_{B}^{0}, c_{C}^{0}) = (0.06, 0.18)$, is steeper and reaches its minimum earlier than the other alloy (in green). The steeper drop in is because of higher supersaturation  $c_{C}^{0}$ for shell growth (i.e., $K$ is higher in Equation~\ref{equation_dsigma_s_dt_multi_ppt_b}); the duration of size focusing $t_\mathrm{sf}$ is shorter due to the smaller inter-precipitate distance, which, in turn, is due to larger number density of precipitates in the alloy with $c_{B}^{0} = 0.06$.

\begin{table}[H]
\centering
    \begin{tabular}{|c|c|ccc|}
    \hline
    $c_{B}^{0}$
        & d    
        & $t_\mathrm{sf}$ 
        & $\sigma^s_\mathrm{min}$
        & $s_\mathrm{sf}$ \\
    \hline
        0.04 & 160 & 1800 & 0.119 & 0.08 \\
        0.06 & 90 & 400  & 0.122 & 0.07 \\
    \hline    
        0.06 & 164 & 2100 & 0.133 & 0.121 \\
    \hline    
    \end{tabular}
    \caption{Metrics of size focusing estimates ($t_\mathrm{sf}$, $\sigma^s_\mathrm{min}$ and $s_\mathrm{sf}$) for the two alloys shown in Figure~\ref{figure_size_focusing_core_vol_fr}; inter-precipitate distance is defined operationally as $d = n^{-1/2}$ where $n$ is the number density of the precipitates.}
    \label{table_size_focusing_core_vol_fr}
\end{table}

In order to clarify the role of inter-precipitate distance, we let the first heat treatment for the alloy with $c_{B}^{0} = 0.06$ proceed to yield a microstructure with a similar inter-precipitate distance ($d \sim 164$) as in the alloy with $c_{B}^{0} = 0.04$ (160); however, this yielded a microstructure with a higher $\sigma^{s}$ and a higher average size ($\overline{r}$ = 20) in the alloy with $c_{B}^{0} = 0.06$. Figure~\ref{figure_size_focusing_core_vol_fr}~(b) shows the comparison between this alloy (in red) and the base alloy $c_{B}^{0} = 0.04$ (in green). The green curve is the same as that in Figure~\ref{figure_size_focusing_core_vol_fr}~(a), whereas the red curve is shifted up due to a microstructure with a wider size distribution. As expected, the difference in the duration of size focusing that we found in Figure~\ref{figure_size_focusing_core_vol_fr}~(a) has now disappeared: $t_\mathrm{sf}$ is nearly the same in the two alloys. Even though the $\sigma^s_\mathrm{min}$ in the red curve is higher, starting $\sigma^s$ is even higher; therefore, the strength of size focusing reported in Table~\ref{table_size_focusing_core_vol_fr} is higher.

\subsection{Role of core coarsening}
\label{section_multi_ppt_core_coarsening}

\begin{figure}[H]
\centering
\includegraphics{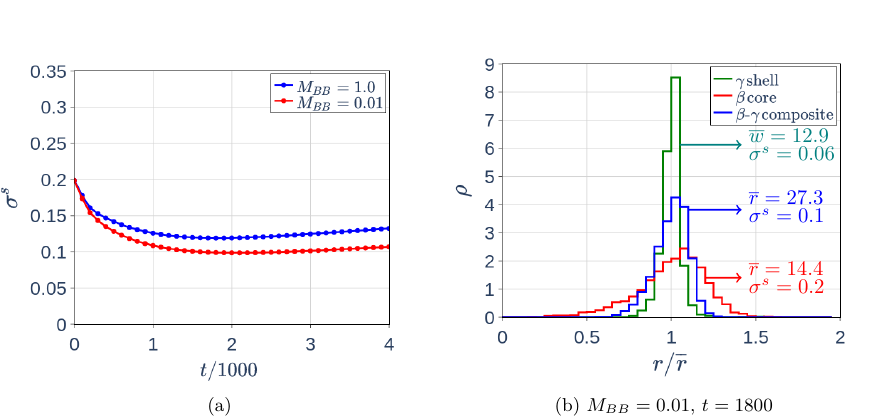}
    \caption{(a) Time evolution of scaled standard deviation ($\sigma^s$) for all precipitates for the alloy with $(c_{B}^{0}, c_{C}^{0}) = (0.04, 0.12)$ for two systems with different core mobilities $M_{BB}$ but with the same shell mobility $M_{CC} = 1$. (b) scaled size distributions for core, shell, and composite precipitates in the system with  $M_{BB} = 0.01$ at $t = 1800$; average radii (widths) and $\sigma^s$ for each phase are shown. [For comparision, see also Figure~\ref{figure_size_focusing_multi_ppt_std_dev_core_shell_rad_size_dist}(c) for $M_{BB} = 1$]}
    \label{figure_size_focusing_multi_ppt_core_coarsening}
\end{figure}

In this section, we compare the size focusing behaviour in the same alloy $(c_{B}^{0}, c_{C}^{0}) = (0.04, 0.12)$ in two systems, one with appreciable ($M_{BB} = 1$) in Figure~\ref{figure_size_focusing_multi_ppt_std_dev_core_shell_rad_size_dist}(c) and another with negligible core mobility ($M_{BB} = 0.01$) in Figure~\ref{figure_size_focusing_multi_ppt_core_coarsening}(a); $\sigma^s$ shows a steeper fall to $\sigma^s_{\text{min}} = 0.1$ in the system with $M_{BB} = 0.01$ compared to $\sigma^s_{\text{min}} = 0.12$ in the system with $M_{BB} = 1$. 

This trend may be explained as follows: while shell growth has a size focusing effect, core coarsening has the opposite (defocusing) effect. But the later (core coarsening) has two components: (a) smaller-than-average core precipitates have shrunk in size and (b) the smallest of them even dissolved away ($\approx$ 5 \%), leaving behind small isolated shells. The net result is that $\sigma^s$ for core radius (0.21), shell width (0.11) and composite radius (0.12) is marigially higher in the system with $M_{BB} = 1$ compared to their values (0.2, 0.06 and 0.1 respectively) in the system with $M_{BB} = 0.01$ (see Figure~\ref{figure_size_focusing_multi_ppt_core_coarsening}(b)); of these, the increase in  $\sigma^s$ for shell width (primarily due to the isolated shell precipitates) is the significant contributer to the increase in overall $\sigma^s$. Therefore, it is clear that suppression of the core coarsening yields a stronger size focusing (i.e., $\sigma^s_{\text{min}}$ is lower; see Table~\ref{table_size_focusing_multi_ppt_core_coarsening}) in the system with $M_{BB} = 0.01$. [These results are broadly in line with two-precipitate simulations presented in Section~C.4 in the Supplementary Material.]

\begin{table}[H]
\centering
    \begin{tabular}{|c|ccc|}
    \hline
    $M_{BB}$
        & $t_\mathrm{sf}$ 
        & $\sigma^s_\mathrm{min}$
        & $s_\mathrm{sf}$ \\
    \hline
        0.01 & 2100 & 0.098 & 0.1 \\
        1.0  & 1800 & 0.119 & 0.08 \\
    \hline    
    \end{tabular}
    \caption{Metrics of size focusing estimates ($\sigma^s_\mathrm{min}$ and $s_\mathrm{sf}$) for the systems shown in Figure~\ref{figure_size_focusing_multi_ppt_core_coarsening}; the differences in $t_\mathrm{sf}$ are not significant.}
    \label{table_size_focusing_multi_ppt_core_coarsening}
\end{table}

\section{Discussion}
\label{section_discussion}

\subsection{Size focusing in binary alloys}
We begin with an important note that a core-shell morphology is not necessary to observe size focusing; it can be observed in a binary alloy. Consider, for example, a binary alloy in which a high-temperature aging produces a small volume fraction of $\beta$ precipitates, followed by aging at a lower temperature that yields a higher volume fraction. If no new $\beta$ nuclei are formed during the second step, further precipitation is through growth over the $\beta$ precipitates inherited from the first step, much like the growth of the shell phase in Section~\ref{section_multi_ppt_alloy_cb0_4_cc0_12}. 

\begin{figure}[H]
\centering
\includegraphics[scale = 0.5]{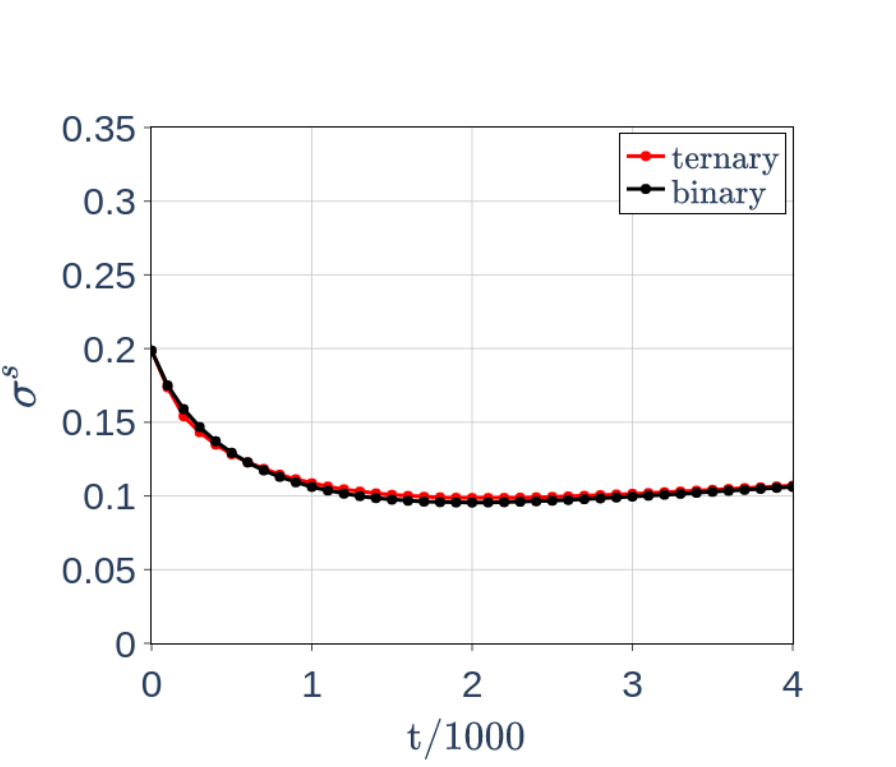}
    \caption{Time evolution of scaled standard deviation ($\sigma^s$) for the binary alloy with $c_{B}^{0} = 0.16$ and its equivalent ternary alloy with $(c_{B}^{0}, c_{C}^{0}) = (0.04, 0.12)$.}
    \label{figure_size_focusing_binary_ternary_size_focusing}
\end{figure}

While such a possibility has not been experimentally attempted so far, we demonstrate it through simulations of a two-step treatment similar to the one we used for the ternary alloy in Section~\ref{section_multi_ppt_alloy_cb0_4_cc0_12}, but with a key difference: after an identical first step (simulation), the second step is simulated with all the shell species ($C$) in the matrix replaced by the solute $B$. [In other words, instead of simulating the second step at a lower temperature (to create additional supersaturation), we just simulate the second step directly with additional supersaturation; this is equivalent to solute-injection in chemical synthesis of nanoparticles.] In Figure~\ref{figure_size_focusing_binary_ternary_size_focusing}, we plot the results for the ternary alloy with $(c_{B}^{0}, c_{C}^{0}) = (0.04, 0.12)$ along with those for its binary equivalent, in which growth of the $\beta$ phase during the second step increases the precipitate volume fraction from 0.04 to 0.16. As expected, the plots of $\sigma^s$ \textit{vs.} time are essentially the same for these two cases.

\subsection{A critique of the use of the LSW theory for studying size focusing}
\label{section_LSW_critique}

In the literature on size focusing, the LSW theory of coarsening forms the basis of several theoretical studies; see, for example, Sugimoto \cite{sugimoto1987preparation}, Clark et al. \cite{clark2011focusing} and the review by Yin and Alivisatos \cite{yin2005colloidal} (for wet chemical synthesis of nanoparticles) as well as Radmilovic et al. \cite{radmilovic2011highly} (for Al-Sc-Li alloys). We begin our critique of this approach by considering the following key result for the growth rate $\left(\dfrac{dr}{dt}\right)$ of a precipitate of size $r$ from the LSW theory \cite{lifshitz1961kinetics, wagner1961theorie}:
\begin{equation}
\frac{dr}{dt} = \frac{K^{LSW}}{r} \left( \frac{1}{r_c} - \frac{1}{r}\right) \thickspace,
\label{equation_dr_dt_vs_r_LSW}
\end{equation}
where $r_c$ is the critical radius below which the precipitates dissolve and above which the precipitates grow. For $r \gg r_c$, we get the following approximation: 
\begin{equation}
\frac{dr}{dt} \approx \frac{K^{LSW}}{r_c \thinspace r} = \dfrac{K_g^{LSW}}{r} \thinspace,
\label{equation_dr_dt_vs_r_LSW_r_much_greater_than_rc}
\end{equation}
where $K_g^{LSW} = \dfrac{K^{LSW}}{r_c}$. 
We first note that the LSW theory is applicable to coarsening, and therefore, it is not appropriate, even in principle, to use it for explaining size focusing, a phenomenon confined only to the growth stage. Though it may be argued that the growth rate expression in Equation~\ref{equation_dr_dt_vs_r_LSW_r_much_greater_than_rc} is similar to that obtained from Zener's theory in Equation~\ref{equation_dr_dt_vs_r_Zener}, it cannot form the basis for a quantitative theory (which we presented in Section~\ref{section_theory_of_size_focusing}). A key reason for this is that the LSW theory is built on the assumption that composition gradients are negligibly small; this assumption (which is valid for the coarsening regime) is clearly not valid during precipitate growth (for which Zener's theory is the most appropriate).

\subsection{Comparison with experimental results}

Since our study is based on simulations of core-shell precipitates in 2D systems, several factors must be considered for a critical comparison of our results with experimental results on Al-Sc-Li alloys \cite{radmilovic2008monodisperse, radmilovic2011highly}.

In the alloy with $(c_{B}^{0}, c_{C}^{0}) \thinspace = \thinspace (0.04, 0.12)$, the ratio of volume fractions of the shell and core phases is 3; in the 2D simulations in our study, this ratio translates to nearly equal values of core radius and shell width. Thus, this alloy is similar to that in the experimental study by Radmilovic et al. \cite{radmilovic2011highly}, where the average core radius and shell width were, respectively, 9.2 nm and 10.5 nm. Further, for their alloy with 8.46 at. \% Li and 0.11 at. \% Sc, the authors have reported the core and shell compositions of \ce{Al3Li_{0.4}Sc_{0.6}} and \ce{Al3Li} (respectively). Since the equilibrium matrix composition has not been reported, we proceed as follows: assuming that the matrix does not contain any Sc, we first estimate a core volume fraction of 0.007. This value in 3D translates (under assumptions of similar inter-precipitate distance $d$ and similar and symmetric size distributions; see Section~E in Supplementary Material for details) to a value of 0.043 for 2D systems; interestingly, this value is close to $c_{B}^{0} = 0.04$ in the base alloy in our 2D study. 

In a recent paper~\cite{chhotray2026core} on core-shell microstructures in an Al-Yb-Li alloy, Chhotray and Gautam reported a bimodal distribution for the core radius \textit{as well as} the shell width. However, the distribution for the composite (core-shell) precipitates is unimodal, clearly indicating that the core radius and shell width are inversely correlated, exactly as expected during the growth of the shell phase; see Figure~\ref{figure_size_focusing_multi_ppt_std_dev_core_shell_rad_size_dist}(b). 

\section{Conclusions}
\label{section_conclusions}

We have presented a theory for the phenomenon of size focusing in alloys with core-shell precipitates in ternary alloys; we have also used phase field simulations for a detailed study of size focusing. Our conclusions are summarised as follows:

\begin{itemize}
    \item Using both theory and (two-precipitate and multi-precipitate) simulations, we have established conclusively that size focusing is a growth phenomenon; i.e., it ends when growth ends. More importantly, size \textit{defocusing} begins when coarsening sets in. 
    \item Size focusing is stronger in systems with a larger shell-to-core volume fraction ratio and a lower core-to-shell mobility ratio.
    \item  The duration of size focusing, on the other hand, depends entirely on the inter-precipitate separation.
    \item Size focusing is also possible in binary systems using a process that mimics repeated solute injection used in the chemical synthesis of nanoparticles.
    \item Our results are in agreement with those from the experimental study by Radmilovic et al. \cite{radmilovic2011highly}.
\end{itemize}

\section{Author contributions}
\textbf{Soumya Mishra}: Conceptualisation, Coding and Simulations, Formal analysis, Validation, Visualisation, Writing – original draft, Writing – review \& editing. 

\textbf{T. A. Abinandanan}: Conceptualisation, Supervision, Writing – review \& editing.

\section{Declaration of competing interest}
The authors declare that they have no known competing financial interests or personal relationships that could have influenced the work reported in this paper.

\section{Acknowlegments}
We gratefully acknowledge the University Grants Commission for funding the early parts of this research through its Networking Resource Centre for Materials program. We thank Abhik Choudhary, Surendra Kumar Makineni and Saswata Bhattacharyya for fruitful discussions. We also thank the anonymous reviewers (especially Reviewer \#2) for their insightful comments on an earlier version of this paper.

\bibliographystyle{elsarticle-num} 
\bibliography{references_size_focusing}

% Redefining appendix section numbering
\renewcommand{\appendixname}{}
\renewcommand{\thesection}{\Alph{section}}

% Restarting Figure number with Section and naming with respect to Section in Appendix
\renewcommand{\thefigure}{\thesection\arabic{figure}}
\makeatletter
\@addtoreset{figure}{section}
\makeatother

% Restarting Table number with Section and naming with respect to Section in Appendix
\renewcommand{\thetable}{\thesubsection\arabic{table}}
\makeatletter
\@addtoreset{table}{subsection}
\makeatother

\appendix

% Resetting table counter at each subsection
\counterwithin{table}{subsection} 
\renewcommand{\thetable}{\thesubsection} % Same as subsection number

\begin{comment}
\newpage
\hspace{0pt}
\vfill
\section*{\centering \huge Supplementary Material}
\label{supplementary_material}
(\textbf{Note to the editor:} Because of the limitation of the template, ``Supplementary Material'' is presented in the same file as the main text. Everything that follows will appear online as ``Supplementary Material'' and is not a part of the main text.)
\vfill
\hspace{0pt}  
\end{comment}

% No page number
\newpage
\thispagestyle{empty} 
\hspace{0pt}
\vfill
\section*{\centering \huge Supplementary Material}
\hspace{5pt}
\section*{\centering \LARGE Size focusing in core-shell precipitates: A phase-field study}
\hspace{1pt}
\section*{\centering \large Soumya Mishra and T. A. Abinandanan} 
\section*{\centering  \normalfont
        Department of Materials Engineering \\
        Indian Institute of Science, Bengaluru - 560012, India.}

\vfill
\hspace{0pt}  

% Restting page number and beginning with a new page
\newpage
\pagenumbering{arabic}
\setcounter{page}{1}
   
\section{Composition profiles during shell growth (second heat treatment)}
\label{section_supplementary_composition_profiles}

\begin{figure}[H]
\centering
\includegraphics{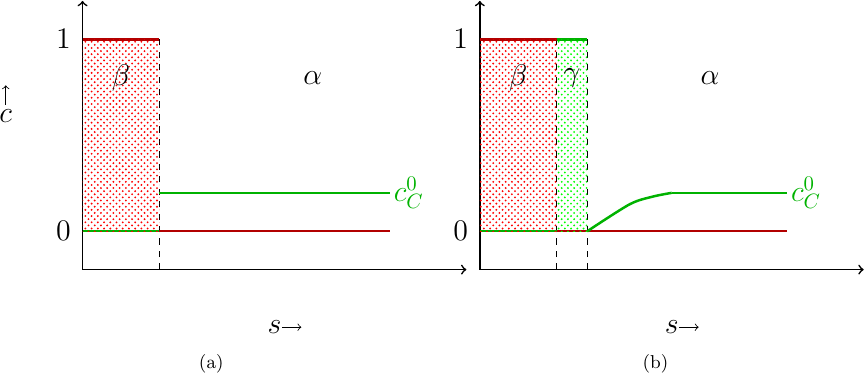}
    \caption{Composition profiles (a) at the beginning of the second heat treatment and (b) after the formation of a thin shell. [The composition variables $c_{B}$ (in red) and $c_{C}$ (in green) are scaled compositions used in our phase field model; in this scheme, $c_{B}^{0}$ and $c_{C}^{0}$ (i.e., the overall alloy composition) are numerically equal to the volume fractions of the core ($\beta$) and shell ($\gamma$) phases, respectively.] After the first treatment, $c_{B} = 1$ within the core ($\beta$) precipitate, and nearly zero in the matrix. Clearly, during the shell phase, it is growing into a matrix with a constant far-field supersaturation $c_{C}^{0}$.}
    \label{figure_supplementary_composition_profiles}
\end{figure}

\section{Theory of size focusing}
\label{section_supplementary_theory_of_size_focusing}

For an ensemble of $n$ precipitates, the average radius ($\overline{r}$) and variance ($\sigma^2$) are given by:
\begin{subequations}
    \begin{align}
        \overline{r} & = \dfrac{\sum_{i=1}^ {n} r_{i}}{n} 
        \label{equation_A1_a}\\
        \sigma^{2} & = \dfrac{\sum_{i=1}^{n} (r_i - \overline{r})^2}{n} =  \overline{r^2} -  \overline{r}^2
          \label{equation_A1}
    \end{align}
\end{subequations}
where $\overline{r^{2}} = (1/n) \sum_{i=1}^ {n} r_{i}^{2}$.
Differentiating $\overline{r}$ and $\sigma^{2}$ with respect to time, and substituting Equation~5, we have:
\begin{equation}
    \begin{aligned}
        \dfrac{d\overline{r}}{dt} & = \dfrac{\sum_{i=1}^ {n} \dfrac{K}{2r_{i}}}{n} = \dfrac{K}{2H}\\
        \dfrac{d\sigma^{2}}{dt} & = \dfrac{\sum_{i=1}^{n} 2r_i (\dfrac{K}{2r_i})}{n} - 2 \overline{r}\dfrac{K}{2H} = K (1 - \dfrac{\overline{r}}{H})
        \label{equation_A2}
    \end{aligned}
\end{equation}
where $H$ is the harmonic mean of the ensemble, defined by $1/H = \sum_{i=1}^ {n} \left[ \, 1 / r_{i} \, \right]/n$.
In order to track the rate of narrowing of the size distribution, we use the following definition of $\sigma^{s}$, the scaled standard deviation for an ensemble of $n$ precipitates: 
\begin{equation}
  \sigma^s = \frac{\sigma}{\overline{r}} \thickspace ,
  \label{equation_A3}
\end{equation}
The time dependence of $\sigma^s$ can be obtained by differentiating the square of Equation~\ref{equation_A3} with respect to time and Substituting the values of $d\sigma^2/dt$ and $d\overline{r}/dt$ from Equation~\ref{equation_A2}:
\begin{equation}
\begin{aligned}
\dfrac{d(\sigma^s)^2}{dt} & = \dfrac{\overline{r}^2\dfrac{d\sigma^2}{dt} - \sigma^2\dfrac{d\overline{r}^2}{dt}}{\overline{r}^4} \thinspace \\
\Rightarrow 2\sigma^s \dfrac{d\sigma^s}{dt} & = \dfrac{K  \thinspace \overline{r}^2 (1 - \dfrac{\overline{r}}{H}) - 2\sigma^2\overline{r} \left(\dfrac{K}{2H} \right)}{\overline{r}^4} \thinspace  \\
\Rightarrow \dfrac{d\sigma^s}{dt} & = \dfrac{K}{2 \sigma^s \overline{r}^3 H} (\overline{r} (H - \overline{r}) - \sigma^2)
\label{equation_A4}
\end{aligned}
\end{equation}
Substituting for $\sigma^2 = \overline{r^2} - \overline{r}^2$, Equation~\ref{equation_A4} can be simplified to:
\begin{equation}
    \dfrac{d\sigma^s}{dt} \thinspace = \thinspace \dfrac{-K}{2 \thinspace \sigma^s \thinspace \overline{r}^3 \thinspace H} \left[\overline{r^2} - H \thinspace \overline{r}\right] \thinspace 
\label{equation_A5} 
\end{equation}
or
\begin{equation}
    \dfrac{d\sigma^s}{dt} = \thinspace \thinspace \dfrac{-K}{2 \thinspace \sigma^s \thinspace \overline{r}} \left[\dfrac{1 \thinspace + \thinspace {\sigma^s}^2}{H} - \dfrac{1}{\overline{r}}\right] \thinspace
\label{equation_A6}
\end{equation}

For a system with only two precipitates of radii $r_{l}$ and $r_{s}$ (with $r_{l} > r_{s}$), Equation~\ref{equation_A6} reduces to Equation~8 wherein $\overline{r} = (r_l + r_s) / 2$, $\sigma^s = (r_ l- r_s) / (r_l + r_s)$ and $H = 2 \thinspace r_s \thinspace r_l/(r_s + r_l)$.

\section{Two-precipitate simulations}
\label{section_supplementary_two_precipitate_simulations}

\subsection{Microstructural evolution}
\label{section_supplementary_two_ppt_microstructures}

\begin{figure}[H]
\centering
\includegraphics{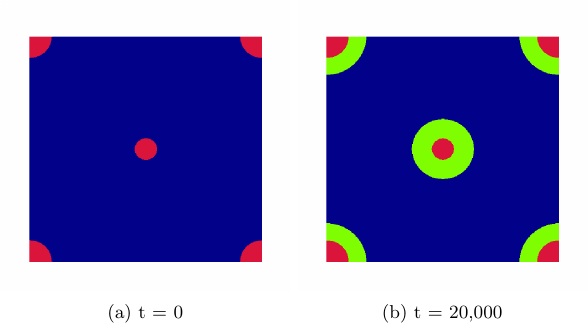}
    \caption{Microstructural evolution (512 x 512) in a two-precipitate simulation for the alloy $(c_{B}^{0}, c_{C}^{0}) = (0.04, 0.12)$. The initial radii are $r_s \approx 24.5$ at the centre and $r_l \approx 49$ at the corner, yielding a radius ratio of $f_{r} = 2$. Mobility of the core-forming species ($M_{BB} = 0.01$) is far lower than that of the shell-forming species ($M_{CC} = 1$); this ensures negligible growth or coarsening of the core precipitates. The shell around the smaller core precipitate is visibly thicker than that around the larger core precipitate.}
    \label{figure_supplementary_two_ppt_microstructures}
\end{figure}

\subsection{Role of shell volume fraction}
\label{section_supplementary_two_ppt_shell_vol_fr}

\begin{figure}[H]
\centering
\includegraphics[scale = 0.5]{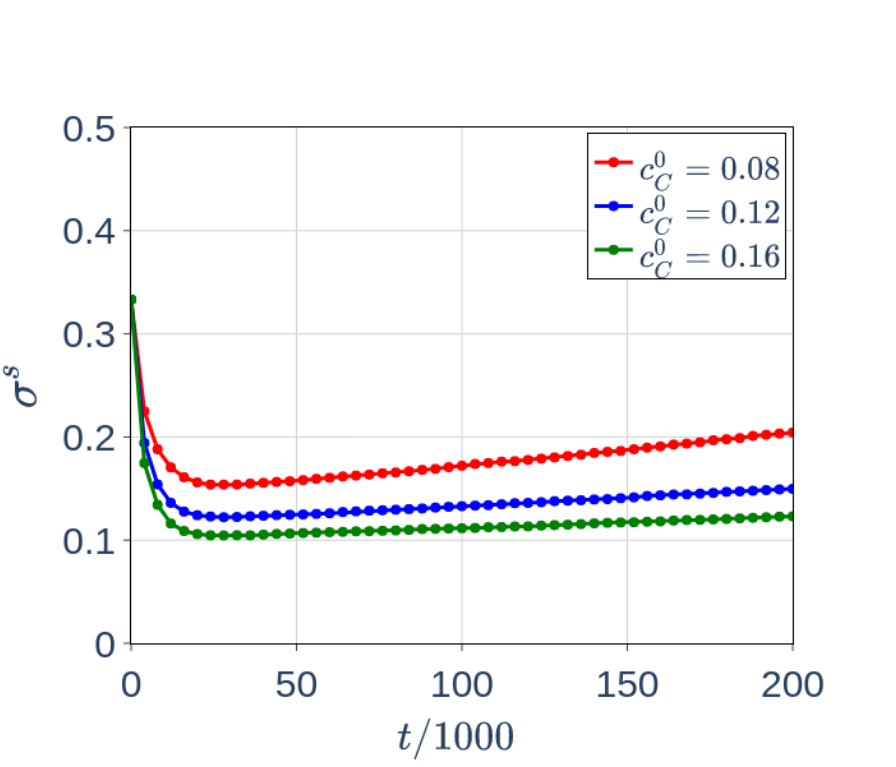}
    \caption{Time evolution of scaled standard deviation ($\sigma^s$) in a two-precipitate simulation for alloys with three shell volume fractions: $c_{C}^{0}$ = 0.08, 0.12, 0.16. The core volume fraction is the same($c_{B}^{0}$ = 0.04) in all three alloys. Mobility of the core-forming species ($M_{BB} = 0.01$) is far lower than that of the shell-forming species ($M_{CC} = 1$); the initial microstructure in all three alloys is the same as Figure~\ref{figure_supplementary_two_ppt_microstructures} (a). With increasing shell volume fraction, size focusing is stronger, and since the inter-precipitate distance is the same, the duration of size focusing is also the same.}
    \label{figure_supplementary_two_ppt_shell_vol_fr}
\end{figure}

\begin{table}[H]
\centering
    \begin{tabular}{|c|c|}
    \hline
    $c_{C}^{0}$ & $s_\mathrm{sf}$ \\
    \hline
        0.08 & 0.177 \\
        0.12 & 0.210 \\
        0.16 & 0.228 \\
    \hline    
    \end{tabular}
    \caption{The strength of size-focusing estimates ($s_{\text{sf}} = \sigma_{0}^{s} - \sigma_{\text{min}}^{s}$) for the three alloys in Figure~\ref{figure_supplementary_two_ppt_shell_vol_fr}}
    \label{table_supplementary_two_ppt_shell_vol_fr}
\end{table}

\subsection{Role of core volume fraction}
\label{section_supplementary_two_ppt_core_vol_fr} 

\begin{figure}[H]
\centering
\includegraphics[scale = 0.5]{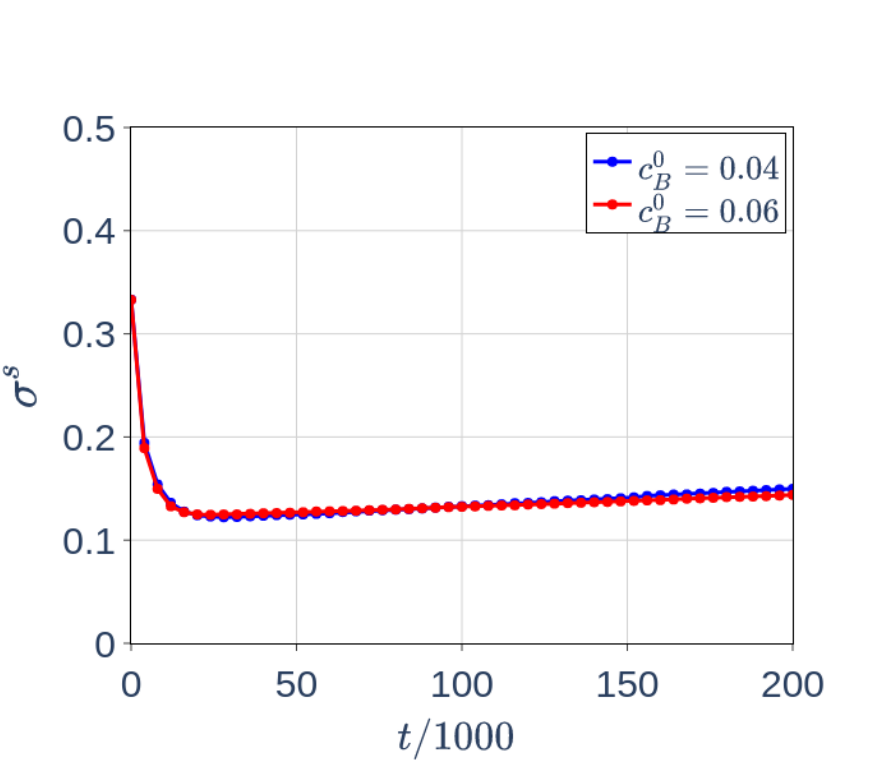}
    \caption{Time evolution of $\sigma^s$ in a two-precipitate simulation for alloys with compositions ($c_{B}^{0}$, $c_{C}^{0}$) = (0.04, 0.12) and (0.06, 0.18) with the same ratio of shell-to-core volume fraction $(c_{C}^{0}/c_{B}^{0}) = 3$; for the second alloy, the starting core sizes are 30 and 60 approximately, yielding a radius ratio of $f_{r} = 2$. Mobility of the core-forming species ($M_{BB} = 0.01$) is far lower than that of the shell-forming species ($M_{CC} = 1$). In the size focusing regime, the behaviours of the two alloys are nearly identical.}
    \label{figure_supplementary_two_ppt_core_vol_fr}
\end{figure}

\begin{table}[H]
\centering
    \begin{tabular}{|c|c|}
    \hline
    $(c_{B}^{0}, c_{C}^{0})$ & $s_\mathrm{sf}$ \\
    \hline
        (0.04, 0.12) & 0.210 \\
        (0.06, 0.18) & 0.208 \\
    \hline    
    \end{tabular}
    \caption{The strength of size-focusing estimates ($s_{\text{sf}}$) for alloys with different core volume fraction ($c_{B}^{0}$) and $(c_{C}^{0}/c_{B}^{0}) = 3$}
    \label{table_supplementary_two_ppt_core_vol_fr}
\end{table}

\subsection{Role of core coarsening}
\label{section_supplementary_two_ppt_core_mobility}

\begin{figure}[H]
\centering
\includegraphics[scale = 0.5]{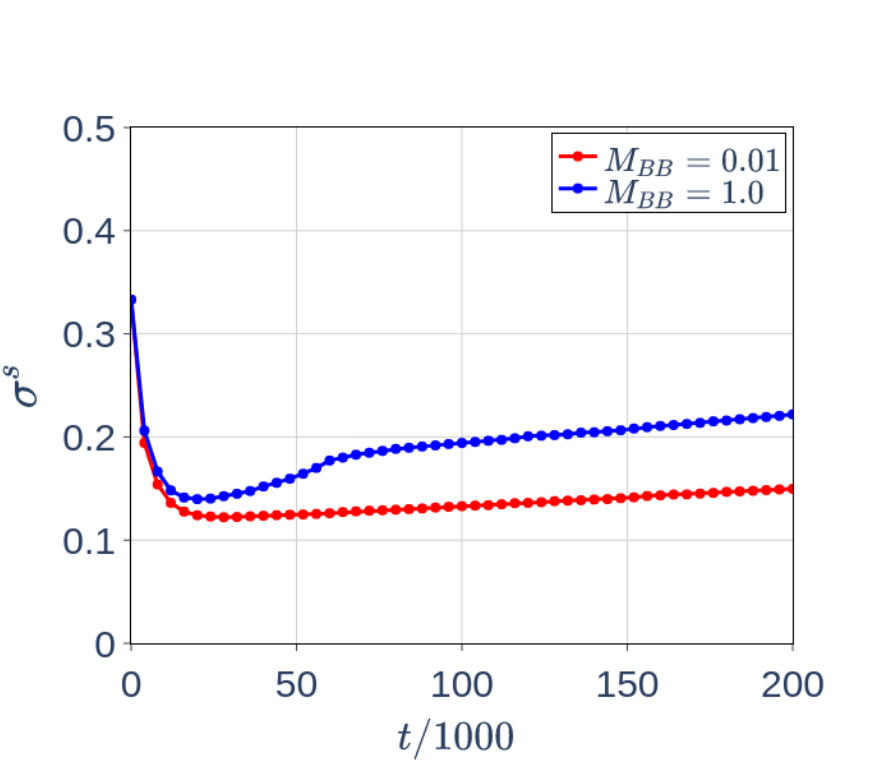}
    \caption{Time evolution of $\sigma^s$ in a two-precipitate simulation for alloy ($c_{B}^{0}$, $c_{C}^{0}$) = (0.04, 0.12) with different mobilities of the core-forming species $M_{BB} = 1$ and $0.01$ ($M_{CC} = 1$ in both the systems). Since the inter-precipitate distance is the same in both systems, the duration of size focusing is also identical. Core coarsening leads to a size defocusing effect in the system with $M_{BB} = 1$; its suppression yields a stronger size focusing in the system with $M_{BB} = 0.01$. In the system with $M_{BB} = 1$, core coarsening has led to the dissolution of the smaller core precipitate by $t = 56,000$.}
    \label{figure_supplementary_two_ppt_core_mobility_std_dev_rad_fr_2}
\end{figure}

\begin{table}[H]
\centering
    \begin{tabular}{|c|c|}
    \hline
    $M_{BB}$ & $s_\mathrm{sf}$ \\
    \hline
        0.01 & 0.228 \\
        1.0  & 0.216 \\
    \hline    
    \end{tabular}
    \caption{The strength of size-focusing estimates ($s_{\text{sf}}$) for alloys with different $M_{BB}$; the last line is for an alloy with radius ratios $f_{r}$ = 3}
    \label{table_supplementary_two_ppt_core_mobility}
\end{table}

\subsection{Role of interfacial energies}
\label{section_supplementary_interfacial_energies}

\begin{figure}[H]
\centering
\includegraphics[scale = 0.5]{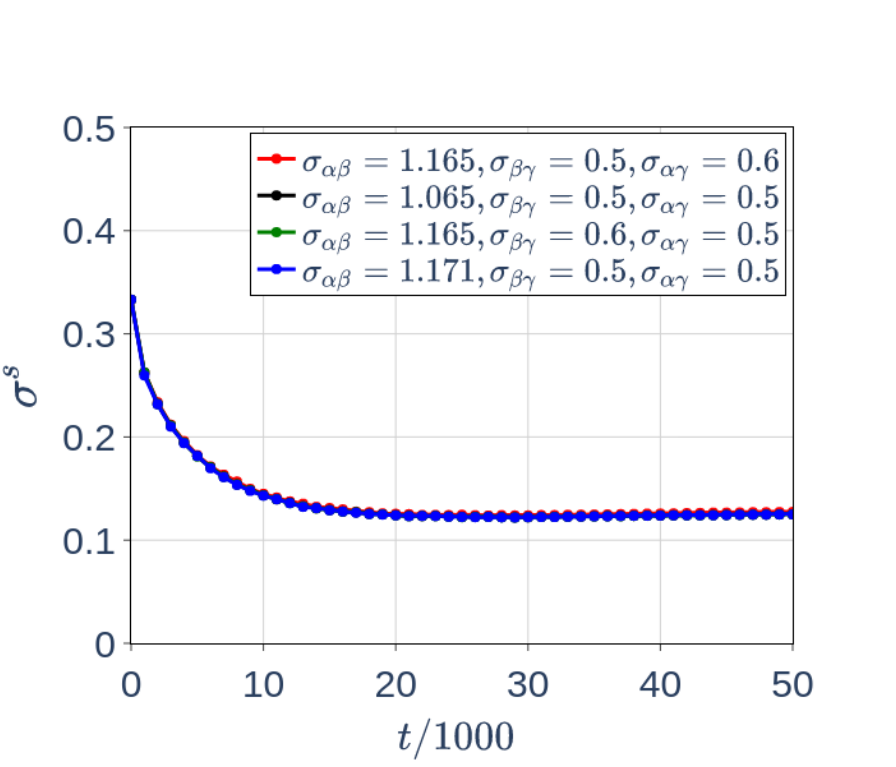}
    \caption{Time evolution of $\sigma^s$ in two-precipitate simulations for alloy with ($c_{B}^{0}$, $c_{C}^{0}$) = (0.04, 0.12) in systems with different interfacial energies. For clarity, the plot is shown only until $t = 50,000$. As expected, interfacial energies have no effect on the strength or duration of size focusing, as long as the wetting condition is satisfied.}
    \label{figure_supplementary_two_ppt_interfacial_energy_std_dev_rad_fr_2}
\end{figure}

\begin{table}[H]
    \centering
    \begin{tabular}{|c|c|c|c|c|c|c|c|}
        \hline
        
        & $(\chi_{AB}$, $\kappa_{AB}$)
        & ($\chi_{BC}$, $\kappa_{BC}$)
        & ($\chi_{AC}$, $\kappa_{AC}$)
        & $\chi_{ABC}$ 
        & $\sigma_{\alpha\beta}$
        & $\sigma_{\beta\gamma}$
        & $\sigma_{\alpha\gamma}$\\ 
        \hline        
        1 & (3, 5.87)   & (1, 2.44) & (1, 3.53) & 50 & 1.165 & 0.5 & 0.6 \\
        2 & (3, 4.88)   & (1, 2.45) & (1, 2.45) & 50 & 1.065 & 0.5 & 0.5 \\
        3 & (3, 5.87)   & (1, 3.53) & (1, 2.44) & 50 & 1.165 & 0.6 & 0.5 \\
        4 & (3.6, 4.88) & (1, 2.45) & (1, 2.45) & 70 & 1.171 & 0.5 & 0.5 \\
        \hline
    \end{tabular} 
    \caption{Model parameters and interfacial energies of four systems shown in Figure~\ref{figure_supplementary_two_ppt_interfacial_energy_std_dev_rad_fr_2}. The four systems differ in interfacial energies, leading to different values of wetting coefficient $S_{\gamma} = \sigma_{\alpha\beta} - (\sigma_{\beta\gamma} + \sigma_{\alpha\gamma})$. In systems 1, 2 and 3, $S_{\gamma} = 0.065$, whereas in System 4, $S_{\gamma} = 0.171$.}
    \label{table_supplementary_two_ppt_interfacial_energy}
\end{table}

\section{Multi-precipitate Simulations}

\subsection{Role of shell volume fraction}
\label{section_supplementary_multi_ppt_shell_vol_fr}

\begin{figure}[H]
\centering
\includegraphics{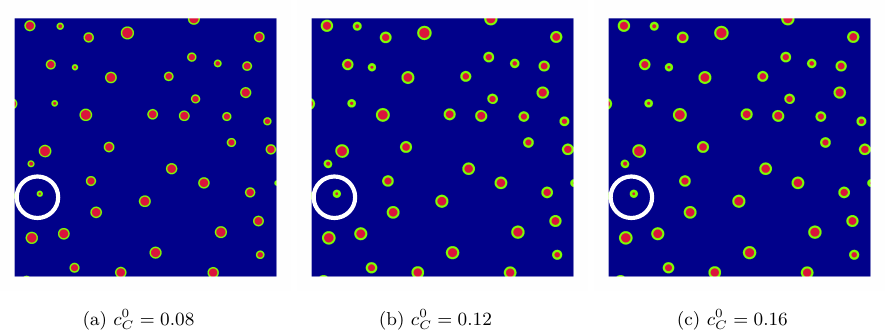}
    \caption{Microstructures (1024 x 1024 section) at t = 800 for three alloys with $c_{C}^{0}$ = 0.08, 0.12, 0.16, with constant $c_{B}^{0}$ = 0.04. The shell width is higher in alloys with higher $c_{C}^{0}$; while this is true for all precipitates, it is more clearly visible in the precipitate highlighted using white circles.}
    \label{figure_supplementary_multi_ppt_shell_vol_fr_microstructures}
\end{figure}

\subsection{Role of core volume fraction}
\label{section_supplementary_multi_ppt_core_vol_fr}

\begin{figure}[H]
\centering
\includegraphics{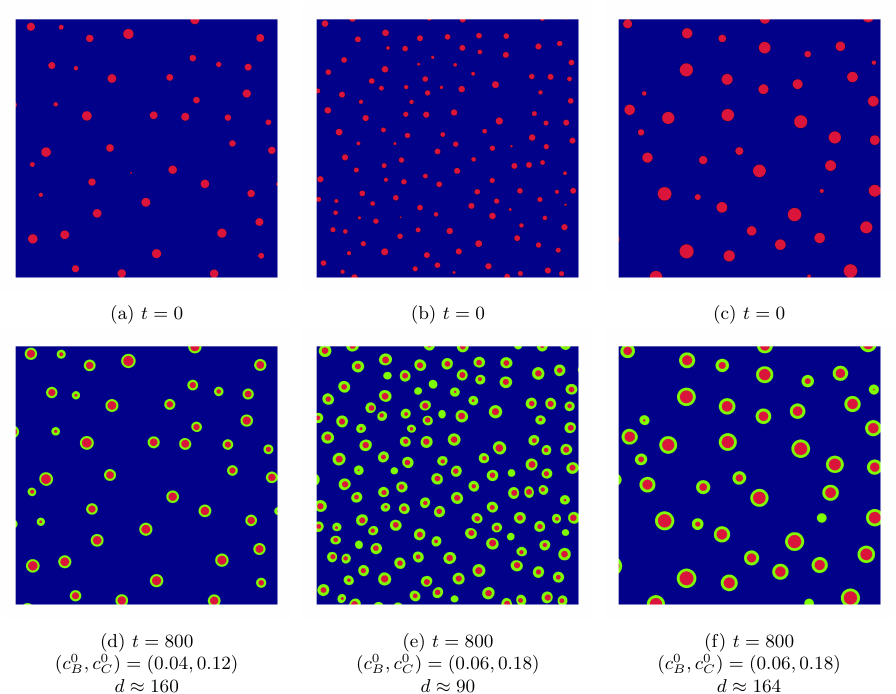}
    \caption{Initial microstructures at the beginning of second heat treatment (t = 0) for (a) $c_{B}^{0} = 0.04$, (b) $c_{B}^{0} = 0.06$ with $c_{C}^{0}$/$c_{B}^{0}$ = 3. These microstructures were prepared by stopping the first simulation when 50 \% of the core precipitates remained. Since the system with $c_{B}^{0} = 0.06$ has a higher core volume fraction, the initial microstructure (b) has a higher density of precipitates. In (c), the first simulation is carried out even further into coarsening so that the inter-precipitate distance is nearly the same as (a). (d), (e) and (f) are the microstructures at $t = 800$ in the respective systems. Since the inter-precipitate distance is smaller in (b) compared to the other two, a significant fraction of core precipitates have dissolved due to faster coarsening in (e), leaving behind isolated shell precipitates in (e) compared to both (d) and (f).}
\label{figure_supplementary_multi_ppt_core_vol_fr_microstructure_and_std_dev_same_d}
\end{figure}

\subsection{Comparision of size focusing in binary and ternary alloys}
\label{section_supplementary_multi_ppt_binary}

\begin{figure}[H]
\centering
\includegraphics{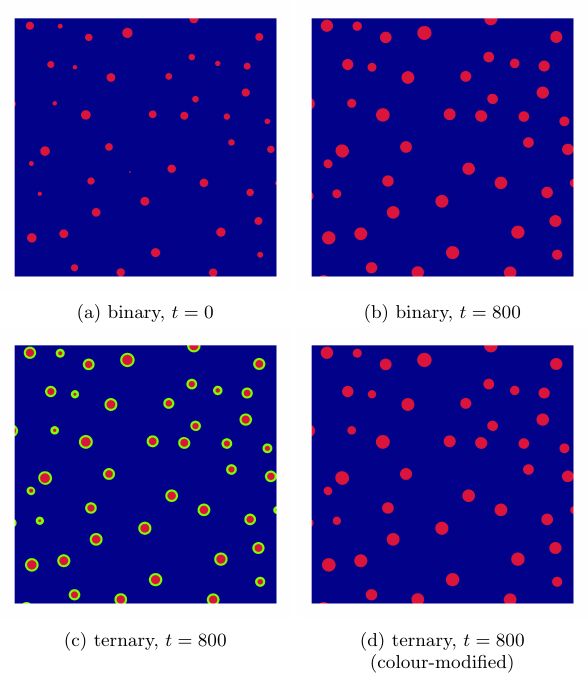}
\caption{Microstructural evolution (1024 x 1024) of binary alloy ($c_{B}^{0} = 0.16$), which is equivalent to the ternary alloy $(c_{B}^{0}, c_{C}^{0}) = (0.04, 0.12)$. (a) and (b) are the microstructures of the binary alloy at $t = 0$ and $t = 800$, respectively. (c) is the microstructure of the equivalent ternary alloy at $t = 800$, whose initial microstructure is the same as (a), and (d) is the same as (c) after removing the distinction between core and shell phases. We note that (b) and (d) are strikingly similar.}
\label{figure_supplementary_binary_microstructures}
\end{figure}

\section{Equivalent volume fractions in 2D and 3D}

\subsection{Equal-sized precipitates}

For an ensemble of $n$ equal-sized precipitates of radius $r$, the volume fraction in 2D is:
\begin{equation}
    v_{2} = \dfrac{n \pi r^2} {A}
    \label{equation_D1}
\end{equation}
where A is the area of the 2D simulation box.
For simpler calculation, if we assume that the precipitates are arranged in a regular square grid with a grid spacing of $d$, then $A = nd^2$. Substituting this in Equation~\ref{equation_D1}, we get:
\begin{equation}
\begin{aligned}
    v_{2} & = \pi \left( \dfrac{r}{d} \right)^2 \\
    \Rightarrow \dfrac{d}{r} & = \sqrt{\dfrac{\pi}{v_{2}}} 
    \label{equation_D2}
\end{aligned}  
\end{equation}
Similarly, for an ensemble of $n$ equal-sized precipitates of radius $r$, the volume fraction in 3D is:
\begin{equation}
    v_{3} = \dfrac{4 n \pi r^3} {3V}
    \label{equation_D3}
\end{equation}
where V is the area of the 3D simulation box.
For simpler calculation, if we assume that $n$ precipitates are arranged in a regular cube grid in 3D with a grid spacing of $d$, then $V = nd^3$. Substituting this in Equation~\ref{equation_D3}:
\begin{equation}
\begin{aligned}
    v_{3} & = \dfrac{4\pi}{3} \left( \dfrac{r}{d} \right)^3 \\
    \Rightarrow \dfrac{d}{r} & = \sqrt[3]{\dfrac{4\pi}{3v_{3}}}
    \label{equation_D4}
\end{aligned}  
\end{equation}
Assuming that the inter-precipitate distance to radius ratio is identical in 2D and 3D (i.e., $(d/r)_{2D} = (d/r)_{3D}$), we can equate Equation~\ref{equation_D2} and Equation~\ref{equation_D4}:
\begin{equation}
    v_2 = \sqrt[3]{\pi}\left( {\dfrac{3v_{3}}{4}} \right)^{2/3}
    \label{equation_D5}
\end{equation}

\subsection{Precipitates with size distribution}
But in real systems, we have a size distribution of precipitates (i.e., finite $\sigma^s$). In such a system, we may derive a relationship between $v_2$ and $v_3$, assuming the following to be similar for both 2D and 3D systems: (a) $d/\overline{r}$, the ratio of inter-precipitate distance to average radius, (b) $\sigma^s$, the scaled standard deviation (c) negligible skewness.

\subsubsection{\texorpdfstring{$d/\overline{r}$}{d/r} in 2D}
The volume fraction of precipitates in 2D ($v_{2}$) can be written as the sum of volumes of all the precipitates:
\begin{equation}
    v_{2} = \dfrac{\pi \sum_{i=1}^{n} r_i^2} {A}
    \label{equation_D6}
\end{equation}
We define, $\overline{r^2}$, the mean square radius of $n$ precipitates with radii $r_i$ as:
\begin{equation}
 \overline{r^2} = \dfrac{\sum_{i=1}^{n} r_i^2}{n} = \sigma_2^2 + \overline{r}^2
 \label{equation_D7}
\end{equation}
where $\overline{r}$ is the average radius and $\sigma_2$ is the standard deviation in 2D.
Using $A = nd^2$ and Equation~\ref{equation_D7}, Equation~\ref{equation_D6} can be rewritten as:
\begin{equation}
    v_{2} = \dfrac{\pi \overline{r^2}} {d^2} = \dfrac{\pi (\sigma_2^2 + \overline{r}^2)} {d^2} = \pi \left( \dfrac{\overline{r}}{d} \right)^2 (1 + {\sigma_2^s}^2)
    \label{equation_D8}
\end{equation}
where $\sigma_2^s = \sigma_2/\overline{r}$ is the scaled standard deviation of precipitates in 2D.
We can rewrite Equation~\ref{equation_D8} as:
\begin{align}
   \dfrac{d}{\overline{r}} & = \sqrt{\dfrac{\pi (1 + {\sigma_2^s}^2)}{v_{2}}} 
   \label{equation_D9}
\end{align}

\subsubsection{\texorpdfstring{$d/\overline{r}$}{d/r} in 3D}

Similarly, the volume fraction of precipitates in 3D ($v_{3}$) can be written either as the sum of volumes of all the precipitates: 
\begin{equation}
    v_{3} = \dfrac{4\pi \sum_{i=1}^{n} r_i^3}{3V}
    \label{equation_D10}
\end{equation}
where V is the volume of the 3D simulation box.
We define, $\overline{r^3}$, the mean cube radius of $n$ precipitates with radii $r_i$ as:
\begin{equation}
 \overline{r^3} = \dfrac{\sum_{i=1}^{n} r_i^3}{n}
 \label{equation_D11}
\end{equation}
The well-known expression for the skewness ($\mu_3$) of an ensemble consisting of $n$ precipitates is given by (see, for example, Chapter 6, pages 6.69-70, in P. N. Arora, \textit{Comprehensive Statistical Methods}, S. Chand Publishing, 2010):
\begin{equation*}
    \mu_3 = \dfrac{\overline{r^3} - 3 \overline{r^2} \overline{r} + 2 \overline{r}^3}{\sigma_3^3}
\end{equation*}
where $\sigma_3$ is the standard deviation of the ensemble in 3D.
If $\mu_3 = 0$, then:
\begin{equation}
\begin{aligned}
    \overline{r^3} & = 3 \overline{r^2} \overline{r} - 2 \overline{r}^3 = 3 (\sigma_3^2 + \overline{r}^2) \overline{r} - 2 \overline{r}^3 \\
\Rightarrow  \overline{r^3} &= 3 \sigma_3^2 \overline{r} + \overline{r}^3
\label{equation_D12}
\end{aligned}
\end{equation}
If we assume that $n$ precipitates are arranged in a regular cube grid in 3D with a grid spacing of $d$, then $V = nd^3$. Using this relation along with Equation~\ref{equation_D11}, Equation~\ref{equation_D10} can be rewritten as:
\begin{equation}
    v_{3} = \dfrac{4\pi \overline{r^3}}{3d^3} = \dfrac{4\pi (3 \sigma_3^2\overline{r}  +\overline{r}^3)}{3d^3} = \dfrac{4 \pi}{3}\left(\dfrac{\overline{r}}{d}\right)^3 (1 + 3 {\sigma_3^s}^2)
    \label{equation_D13}
\end{equation}
where $\sigma_3^s = \sigma_3/\overline{r}$ is the scaled standard deviation in 3D.
We can rewrite Equation~\ref{equation_D13} as:
\begin{align}
   \dfrac{d}{\overline{r}} & = \sqrt[3]{\dfrac{4 \pi (1 + 3{\sigma_3^s}^2)}{3v_{3}}}
    \label{equation_D14}
\end{align}
Assuming that the inter-precipitate distance to average radius ratio is identical in 2D and 3D (i.e., $(d/\overline{r})_{2D} = (d/\overline{r})_{3D}$), we can equate Equation~\ref{equation_D8} and Equation~\ref{equation_D14}:
\begin{equation}
    v_2 = \sqrt[3]{\pi} (1 + {\sigma_2^s}^2) \left( {\dfrac{3v_{3}}{4 (1 + 3{\sigma_3^s}^2)}} \right)^{2/3}
    \label{equation_D15}
\end{equation}
For a system with equal-sized precipitates, $\sigma^s = 0$; in this case Equation~\ref{equation_D15} reduces to Equation~\ref{equation_D5}.

\section{Determination of duration of classical growth}

As discussed in Section~3, our theory of size focusing builds on the Zener result in Equation~5. In a multi-precipitate system, Zener's theory is valid only for the early stages before diffusion fields from neighbouring precipitates overlap. [Afterwards, the growth proceeds in a non-classical mode in an environment of ever-decreasing supersaturation.] In what follows, we use a simple 1D growth model to estimate the duration ($\tau$) during which the growth follows the classical Zener dynamics.

\begin{figure}[H]
\centering
    \includegraphics[scale = 0.75]{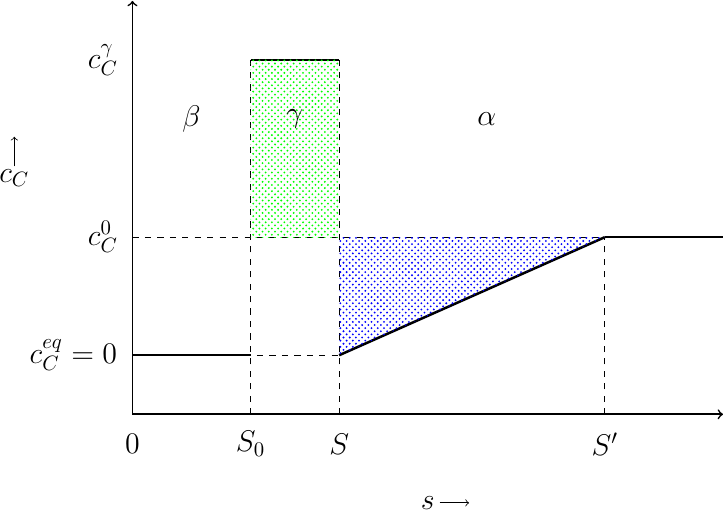}
    \caption{Schematic showing composition profile for 1D growth of the shell phase ($\gamma$) on a pre-existing beta precipitate in the matrix ($\alpha$).}
    \label{figure_sharp_interface_alpha_gamma}
\end{figure}

We consider the planar (1D) growth of the shell phase ($\gamma$) on a pre-existing core ($\beta$) into an infinite matrix ($\alpha$) with constant far-field supersaturation in $C$ ($c_{C}^{0}$). We use a well-known approximation method in which the composition profile (in terms of $c_C$) in the matrix is approximated to a straight line, as shown in the schematic in Figure~\ref{figure_sharp_interface_alpha_gamma}. The core-shell interface is at $S_0$, and the shell-matrix interface is at $S$.

At time $t = 0$ of the second (lower temperature) treatment, the entire matrix outside $\beta$ precipitate was at a composition $c_C^0$ (see Figure~\ref{figure_supplementary_composition_profiles} (a)). Therefore, conservation of solute $C$ in Figure~\ref{figure_sharp_interface_alpha_gamma} implies that the area of the shaded region in the shell phase (in green) must be equal to the shaded region in the matrix phase (in blue):
\begin{subequations}
\begin{align}
( c_{C}^{\gamma} - c_{C}^{0} ) (S - S_0) & = 
    \dfrac{1}{2} ( c_{C}^{0} - c_{C}^{eq} ) ( S^{\prime} - S ) \\
\implies S^{\prime} &  = A + B S, 
\label{equation_E1b}
\end{align}
\end{subequations}
where $A$ and $B$ are constants given by $A = -2 \dfrac{1 - v_{\gamma}}{v_{\gamma}} S_0$ and $B = \dfrac{(2 - v_{\gamma})}{v_{\gamma}}$. [For our phase-field model, $c_{C}^{\gamma} = 1.0$, $c_{C}^{eq} = 0$ and $c_{C}^{0}$ is the volume fraction of shell ($\gamma$) phase ($v_{\gamma}$)].

This precipitate is growing in an environment with another precipitate at $s = d$. Let the two diffusion fields begin to overlap at $t = \tau$. Clearly, at $t = \tau$, $S^{\prime} = d/2$ and for this time, let the shell-matrix interface be at $S_{\tau}$.

From Zener's growth law:
\begin{align}
 (S_{\tau})^2 = S_0^2 + K \tau  
\label{equation_E3}
\end{align}

From Equation~\ref{equation_E1b}, we have:
\begin{subequations}
\begin{align}
 S_{\tau} =  \dfrac{S^{\prime} - A}{B} \\
 \implies S_{\tau} ^2 = \dfrac{1}{B^2} \left( \dfrac{d}{2} - A \right)^2,
\label{equation_E4b}
\end{align}
\end{subequations}
since $S^{\prime} = d/2$.
Equating $S_{\tau} ^2$ in Equations~\ref{equation_E3} and \ref{equation_E4b} and rearranging, we have:
\begin{equation}
    \tau = \dfrac{\dfrac{1}{B^2} \left[ \dfrac{d}{2} - A \right]^2 - S_0^2}{K}
    \label{equation_E5}
\end{equation}

Substituting $d = 362$, $K = 0.26$ (Section~3.1), $v_{\gamma} = 0.12$ and $S_0 = v_{\beta} \thinspace d/2 = 7.24$, we have $\tau \approx 1090$. This estimate of $\tau$ agrees to within a factor of 2.5, with the value obtained in our two-precipitate simulations in Section~3.1.

\section{Colour Palette}

\begin{figure}[H]
\centering
    \includegraphics[scale = 0.5]{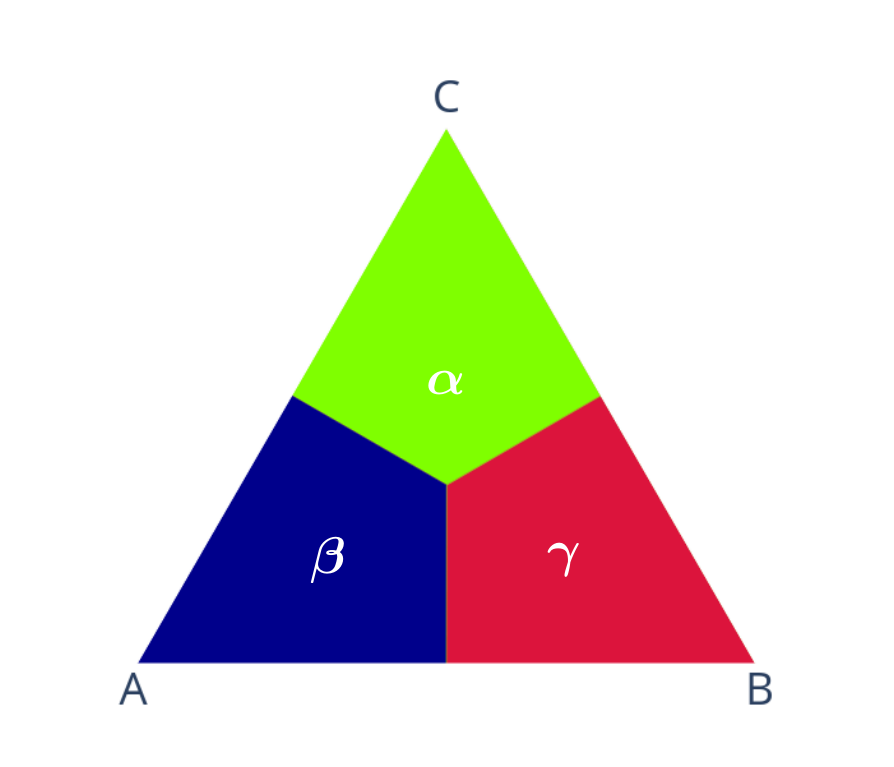}
    \caption{Colour pallette used to represent the phases $\alpha$ (A), $\beta$ (B) and $\gamma$ (C)}
    \label{figure_analysis_of_microstructures_phase_colour}
\end{figure}

\section{Phase Field Model}

The time evolution of the composition of each species $i$ is given by the continuity equation:
\begin{equation}
\frac{\partial c_{i}}{\partial t }= -\nabla \cdot \tilde{J}_{i} \label{equation_3.4}
\end{equation}
where $\tilde{J}_{i}$ is the flux of species $i$ with respect to the laboratory frame of reference (fixed to the one end of the specimen)
The net flux of all three species is zero in the laboratory frame of reference:
\begin{equation}
\sum_{i} \tilde{J}_{i} = 0
\label{equation_3.5}
\end{equation}
The diffusion flux of species $i$ (${J}_{i}$) with respect to the marker plane frame of reference (a frame of reference fixed to the marker plane) is:
\begin{equation}
J_{i} = - M_{i} \nabla \mu_{i} 
\label{equation_3.6}
\end{equation}
where $M_{i}$ is the atomic mobility of solute $i$ and $\mu_{i}$ is the chemical potential of species $i$.
We can write the laboratory frame diffusion flux of species $i$ in terms of marker plane fluxes as:
\begin{equation}
\tilde{J}_{i} = J_{i} - c_i(J_{A} + J_{B} + J_{C})
\label{equation_3.7}
\end{equation}
Substituting Equation~\ref{equation_3.5} in Equation~\ref{equation_3.7} leads to 
\begin{align}
\tilde{J}_{A} & = -\left(1-c_{A}\right) M_{A} \nabla \mu_{A}+c_{A} M_{B} \nabla \mu_{B}+c_{A} M_{C} \nabla \mu_{C} \nonumber \\
\tilde{J}_{B} & = -\left(1-c_{B}\right) M_{B} \nabla \mu_{B}+c_{B} M_{A} \nabla \mu_{A}+c_{B} M_{C} \nabla \mu_{C}  \nonumber \\
\tilde{J}_{C} & = -\left(1-c_{C}\right) M_{C} \nabla \mu_{C}+c_{C} M_{A} \nabla \mu_{A}+c_{C} M_{B} \nabla \mu_{B} 
\label{equation_3.8}
\end{align}
The Gibbs-Duhem equation is:
\begin{equation}
c_{B} \nabla \mu_{B}+c_{C} \nabla \mu_{C}+\left(1-c_{B}-c_{C}\right) \nabla \mu_{A} = 0  
\label{equation_3.9}
\end{equation}
Substituting for $c_{A} = 1 - c_{B} - c_{C}$ in Equation~\ref{equation_3.9} and rearranging, we have:
\begin{align}
\nabla \mu_{A} & = -c_{B} \nabla \mu_{B}^{\mathrm{eff}}-c_{C} \nabla \mu_{C}^{\mathrm{eff}} \nonumber \\
\nabla \mu_{B} & = \left(1-c_{B}\right) \nabla \mu_{B}^{\mathrm{eff}}-c_{C} \nabla \mu_{C}^{\mathrm{eff}} \nonumber \\
\nabla \mu_{C} & = \left(1-c_{C}\right) \nabla \mu_{C}^{\mathrm{eff}}-c_{B} \nabla \mu_{B}^{\mathrm{eff}}
\label{equation_3.10}
\end{align}
where $\mu_{B}^{\mathrm{eff}} = \mu_{B}-\mu_{A}$ and $\mu_{C}^{\mathrm{eff}} = \mu_{C}-\mu_{A}$ are the effective chemical potentials of B and C, respectively. 

Substituting Equations~\ref{equation_3.10} in  Equations~\ref{equation_3.8}, we arrive at the following expressions for $\tilde{J}_B$ and $\tilde{J}_C$:
\begin{align*}
\tilde{J}_{B}= & -\left[\left(1-c_{B}\right)^2 M_{B}+c_{B}^2\left(M_{C}+M_{A}\right)\right] \nabla \mu_{B}^{\mathrm{eff}} \\
& +\left[\left(1-c_{B}\right) c_{C} M_{B}+\left(1-c_{C}\right) c_{B} M_{C}-c_{B} c_{C} M_{A}\right] \nabla \mu_{C}^{\mathrm{eff}} 
\end{align*}
\begin{align*}
\tilde{J}_{C} & =\left[c_{B}\left(1-c_{C}\right) M_{C}+\left(1-c_{B}\right) c_{C} M_{B}-c_{B} c_{C} M_{A}\right] \nabla \mu_{B}^{\mathrm{eff}} \\
& -\left[\left(1-c_{C}\right)^2 M_{C}+c_{C}^2\left(M_{B}+M_{A}\right)\right] \nabla \mu_{C}^{\mathrm{eff}} 
\end{align*}
Rewriting these equations in compact form, we have: 
\begin{align}
& \tilde{J}_{B}=-M_{B B} \nabla \mu_{B}^{\mathrm{eff}}-M_{B C} \nabla \mu_{C}^{\mathrm{eff}} \nonumber \\
& \tilde{J}_{C}=-M_{C C} \nabla \mu_{C}^{\mathrm{eff}}-M_{B C} \nabla \mu_{B}^{\mathrm{eff}}
\label{equation_3.11}
\end{align}
where the effective mobilities are given by: 
\begin{align}
M_{B B} & =\left(1-c_{B}\right)^2 M_{B}+c_{B}^2\left(M_{C}+M_{A}\right) \nonumber \\
M_{C C} & =\left(1-c_{C}\right)^2 M_{C}+c_{C}^2\left(M_{B}+M_{A}\right) \nonumber \\
M_{B C} = M_{C B} & = c_{B} c_{C} M_{A} - \left(1-c_{B}\right) c_{C} M_{B} - c_{B} M_{C}\left(1-c_{C}\right)  
\label{equation_3.12}
\end{align}
Equation~\ref{equation_3.11} can be written in matrix form:
\begin{equation*}
\begin{bmatrix}
\tilde{J}_{B} \\
\tilde{J}_{C}
\end{bmatrix}
= -
\begin{bmatrix}
M_{BB} & M_{BC} \\
M_{CB} & M_{CC} \\
\end{bmatrix}
\begin{bmatrix}
\nabla \mu_{B}^{\mathrm{eff}}\\
\nabla \mu_{C}^{\mathrm{eff}} 
\end{bmatrix} 
\end{equation*}
We note that although the effective mobilities are composition dependent, as shown in Equation~\ref{equation_3.12}, we use mobilities that are independent of local composition. For this purpose, we calculate $M_{BB}$, $M_{CC}$ and $M_{BC}$ at the A-rich corner of the Gibbs triangle using the composition $(c_{A}, c_{B}, c_{C}) = (1, 0, 0)$ such that $M_{BB} = M_{CC} = 1$ and $M_{BC} = 0$.

In a homogeneous ternary solid solution, $\mu_{i}^{\mathrm{eff}}$ is the partial derivative of the bulk free energy $f$ with respect to the composition $c_i$:
\begin{equation*}
\mu_{i}^{\mathrm{eff}}=\frac{\partial f}{\partial c_{i}} 
\end{equation*}
This can be extended to ternary inhomogeneous systems, where $\mu_{i}^{\text {eff }}$ as the variational derivative of $F$ with respect to the composition $c_i$:
\begin{equation}
\mu_{i}^{\mathrm{eff}}=\frac{\delta F}{\delta c_{i}} 
\label{equation_3.13}
\end{equation}
The Euler-Lagrange equation for the variational derivative is:
\begin{equation}
\frac{\delta F}{\delta c_{i}}=\frac{\partial F}{\partial c_{i}}-\nabla \cdot \frac{\partial F}{\partial \nabla c_{i}} 
\label{equation_3.14}
\end{equation}
Therefore, $\mu_{B}^{\mathrm{eff}}$ and $\mu_{C}^{\mathrm{eff}}$ are:
\begin{align}
\mu_{B}^{\mathrm{eff}} & =\frac{\partial f_0}{\partial c_{B}}-2\left(\kappa_{B}+\kappa_{A}\right) \nabla^2 c_{B}-2 \kappa_{A} \nabla^2 c_{C} \nonumber \\
\mu_{C}^{\mathrm{eff}} & =\frac{\partial f_0}{\partial c_{C}}-2\left(\kappa_{A} + \kappa_{C}\right) \nabla^2 c_{C}-2 \kappa_{A} \nabla^2 c_{B} 
\label{equation_3.15}
\end{align}
where $\left(\partial f_0 / \partial c_{B}\right)$ and $\left(\partial f_0 / \partial c_{C}\right)$ are given by:
\begin{align}
\frac{\partial f_0}{\partial c_{B}} & = 2\chi_{AB} c_A c_B(c_A - c_B) + 2c_C^2 (\chi_{BC}c_B - \chi_{AC} c_A) + 2 \chi_{ABC} c_A c_B c_C^2 (c_A - c_B) \nonumber \\
\frac{\partial f_0}{\partial c_{C}} & = 2\chi_{AC} c_A c_C(c_A - c_C) + 2c_B^2 (\chi_{BC}c_C - \chi_{AB} c_A) + 2 \chi_{ABC} c_A c_B^2 c_C (c_A - c_C)
\nonumber
\end{align}
Substituting Equations~\ref{equation_3.11} and~\ref{equation_3.14} in Equation~\ref{equation_3.4} and assuming $M_{BB}, M_{CC}$ and $M_{B C}$ to be constants independent of composition, we have:
\begin{align*}
\frac{\partial c_{B}}{\partial t} & = M_{B B}\left[\nabla^2\left(\frac{\partial f_0}{\partial c_{B}}\right)-2\left(\kappa_{A} + \kappa_{B}\right) \nabla^{4} c_{B}-2 \kappa_{A} \nabla^{4} c_{C}\right] \\
& + M_{B C}\left[\nabla^2\left(\frac{\partial f_0}{\partial c_{C}}\right)-2\left(\kappa_{A} + \kappa_{C}\right) \nabla^{4} c_{C}-2 \kappa_{A} \nabla^{4} c_{B}\right] 
\end{align*}
\begin{align}
\frac{\partial c_{C}}{\partial t} & = M_{C C}\left[\nabla^2\left(\frac{\partial f_0}{\partial c_{C}}\right)-2\left(\kappa_{A} + \kappa_{C}\right) \nabla^{4} c_{C}-2 \kappa_{A} \nabla^{4} c_{B}\right] \nonumber \\
& + M_{B C}\left[\nabla^2\left(\frac{\partial f_0}{\partial c_{B}}\right)-2\left(\kappa_{A} + \kappa_{B}\right) \nabla^{4} c_{B}-2 \kappa_{A} \nabla^{4} c_{C}\right] 
\label{equation_3.16}
\end{align}
Since $\frac{\partial f_0}{\partial c_{B}}$ and $\frac{\partial f_0}{\partial c_{C}}$ have terms non-linear in $c_{i}$, these time evolution equations are non-linear and have to be solved numerically. 

\section{Numerical implementation of Ternary Cahn--Hilliard Equations in Fourier Space}

We describe a semi-implicit method for solving the ternary Cahn-Hilliard equation. We have used the CUDA programming language to parallelise the solution of the semi-implicit difference equation using GPU.

We begin with the Fourier transform of Equations~\ref{equation_3.16} (which is the same as Equation~4 in Section~2):
\begin{align}
\frac{\partial \tilde{c}_{B}(\mathbf{k}, t)}{\partial t} & =M_{B B}\left[-k^2 \tilde{g}_{B}(\mathbf{k})-2 \kappa_{AB} k^{4} \tilde{c}_{B}-2 \kappa_{A} k^{4} \tilde{c}_{C}\right] \nonumber \nonumber \\
& -M_{B C}\left[-k^2 \tilde{g}_{C}(\mathbf{k})-2 \kappa_{A} k^{4} \tilde{c}_{B}-2 \kappa_{AC} k^{4} \tilde{c}_{C}\right] \nonumber \\
\frac{\partial \tilde{c}_{C}(\mathbf{k}, t)}{\partial t} & =M_{C C}\left[-k^2 \tilde{g}_{C}(\mathbf{k})-2 \kappa_{AC} k^{4} \tilde{c}_{C}-2 \kappa_{A} k^{4} \tilde{c}_{B}\right] \nonumber \\
& -M_{B C}\left[-k^2 \tilde{g}_{B}(\mathbf{k})-2 \kappa_{A} k^{4} \tilde{c}_{C}-2 \kappa_{AB} k^{4} \tilde{c}_{B}\right] 
\label{equation_E17}
\end{align}
where $\tilde{c}_{B}(\mathbf{k}, t)$ and $\tilde{c}_{C}(\mathbf{k}, t)$ are the Fourier transforms of the compositions in real space, $g_{B}=\frac{\partial f_0}{\partial c_{B}}$,  $g_{C}=\frac{\partial f_0}{\partial c_{C}}$ and $\mathbf{k}$ is a vector in Fourier space and $k=|\mathbf{k}|$ and $\kappa_{A B}=\kappa_{A} + \kappa_{B}$ and $\kappa_{A C} = \kappa_{C} + \kappa_{C}$.

Now, in Equation~\ref{equation_E17}, treating the linear terms $\tilde{c}_{B}$ and $\tilde{c}_{C}$ implicitly (using their values at $(t+\Delta t)$), and the non-linear terms $\tilde{g}_{B}$ and $\tilde{g}_{C}$ explicitly (using their values at $t$), Equations~\ref{equation_E17} can be written in a compact form:
\begin{equation}
\begin{bmatrix}
P_{1} & P_2  \\
P_{3} & P_{4}
\end{bmatrix}
\begin{bmatrix}
\tilde{c}_{B}(t+\Delta t) \\
\tilde{c}_{C}(t+\Delta t)
\end{bmatrix}
=
\begin{bmatrix}
Q_{1} \\
Q_2
\end{bmatrix}
\label{equation_E27}\\
\end{equation}
where
\begin{align*}
P_{1} & = 1+2 M_{B B} \thinspace \kappa_{AB} k^{4} \Delta t-2 M_{B C} \thinspace \kappa_{A} k^{4} \Delta t \\
P_2 & = 2\left(M_{B B} \thinspace \kappa_{A}-M_{B C} \thinspace \kappa_{AC}\right) k^{4} \Delta t \\
\end{align*}
\begin{align*}
P_{3} & = 2\left(M_{C C} \thinspace \kappa_{A}-M_{B C} \thinspace \kappa_{AB}\right) k^{4} \Delta t \\
P_{4} & = 1+2 M_{C C} \thinspace \kappa_{AC} k^{4} \Delta t-2 M_{B C} \thinspace \kappa_{A} k^{4} \Delta t  \\
Q_{1} & = \tilde{c}_{B}(t)-M_{B B} \thinspace k^2 \frac{\partial f}{\partial c_{B}} \Delta t+M_{B C} \thinspace k^2 \frac{\partial f}{\partial c_{C}} \Delta t \\
Q_2 & = \tilde{c}_{C}(t)-M_{C C} \thinspace k^2 \frac{\partial f}{\partial c_{C}} \Delta t+M_{B C} \thinspace k^2 \frac{\partial f}{\partial c_{B}} \Delta t 
\end{align*}

Therefore, if we know the composition fields at time $t$, we can numerically solve the above equations to find composition fields at $t + \Delta t$; repeating this process over and over again in finite steps of size $\Delta t$ allows us to study the microstructure. 

\section{Non-Dimensionalisation of the Ternary Cahn--Hilliard System}

The scaling and non-dimensionalisation procedure is the same as that used by  Gururajan and Abinandanan [Phase ﬁeld study of precipitate rafting under a uniaxial stress, Acta Materialia 55 (15) (2007) 5015–5026] for the binary Cahn-Hilliard Equation.quation.

The rescaling of the compositions ($c_B$ and $c_C$) is done in such a way that they are the same as the volume fractions of $\beta$ and $\gamma$ precipitates, respectively. All equations used in our model are non-dimensionalised by scaling with the following quantities: characteristic energy $E$, characteristic length $L$, and characteristic time $T$. For example, the non-dimensional counterparts of $N^{\prime}_v$, $f^{\prime}_0$,  $\kappa^{\prime}$ and $M^{\prime}$ can be obtained as:
\begin{align}
    N_v      & = N^{\prime}_v \ L^3,                  \\
    f_0      & = \frac{f^{\prime}_0}{E}               \\
    \kappa_i & = \frac{\kappa^{\prime}_i}{E L^2}      \\
    M_{CC}   & = \frac{T M_{CC}^{\prime} E}{L^2}
\end{align}
where $N^{\prime}_v$ is the number of atoms per unit volume ($\text{atoms/m}^3$),$f^{\prime}_0$ is the bulk free energy per atom ($\text{J/atom}$), $\kappa^{\prime}_i$ is the gradient energy coefficient for species $i$ ($\text{J m}^2/\text{atom}$) and $M_{CC}^{\prime}$ is the effective mobility of species $C$  ($\text{m}^2/\text{Js}$). [We note that we use $M_{CC}$ for non-dimensionalisation as  $M_{CC} = 1$ across all simulations in our study.]

\end{document}